# 10. ELECTROWEAK MODEL AND CONSTRAINTS ON NEW PHYSICS

Revised December 2003 by J. Erler (U. Mexico) and P. Langacker (Univ. of Pennsylvania).

- 10.1  Introduction
- 10.2  Renormalization and radiative corrections
- 10.3  Cross-section and asymmetry formulas
- 10.4  $W$ and $Z$ decays
- 10.5  Experimental results
- 10.6  Constraints on new physics

## 10.1. Introduction

The standard electroweak model (SM) is based on the gauge group [1] SU(2) × U(1), with gauge bosons $W^i_\mu$, $i = 1,2,3$, and $B_\mu$ for the SU(2) and U(1) factors, respectively, and the corresponding gauge coupling constants $g$ and $g'$. The left-handed fermion fields $\psi_i = \begin{pmatrix} \nu_i \\ \ell_i^- \end{pmatrix}$ and $\begin{pmatrix} u_i \\ d'_i \end{pmatrix}$ of the $i^{th}$ fermion family transform as doublets under SU(2), where $d'_i \equiv \sum_j V_{ij} d_j$, and $V$ is the Cabibbo-Kobayashi-Maskawa mixing matrix. (Constraints on $V$ and tests of universality are discussed in Ref. 2 and in the Section on the Cabibbo-Kobayashi-Maskawa mixing matrix.) The right-handed fields are SU(2) singlets. In the minimal model there are three fermion families and a single complex Higgs doublet $\phi \equiv \begin{pmatrix} \phi^+ \\ \phi^0 \end{pmatrix}$.

After spontaneous symmetry breaking the Lagrangian for the fermion fields is

$$\begin{aligned}
\mathscr{L}_F &= \sum_i \overline{\psi}_i \left( i\,\slashed{\partial} - m_i - \frac{g m_i H}{2 M_W} \right) \psi_i \\
&\quad - \frac{g}{2\sqrt{2}} \sum_i \overline{\psi}_i \gamma^\mu \left(1 - \gamma^5\right) \left(T^+ W^+_\mu + T^- W^-_\mu\right) \psi_i \\
&\quad - e \sum_i q_i \,\overline{\psi}_i \gamma^\mu \psi_i\, A_\mu \\
&\quad - \frac{g}{2\cos\theta_W} \sum_i \overline{\psi}_i \gamma^\mu \left(g^i_V - g^i_A \gamma^5\right) \psi_i\, Z_\mu \,. \qquad (10.1)
\end{aligned}$$

$\theta_W \equiv \tan^{-1}(g'/g)$ is the weak angle; $e = g\sin\theta_W$ is the positron electric charge; and $A \equiv B\cos\theta_W + W^3 \sin\theta_W$ is the (massless) photon field. $W^\pm \equiv (W^1 \mp iW^2)/\sqrt{2}$ and $Z \equiv -B\sin\theta_W + W^3 \cos\theta_W$ are the massive charged and neutral weak boson fields, respectively. $T^+$ and $T^-$ are the weak isospin raising and lowering operators. The vector and axial-vector couplings are

$$g^i_V \equiv t_{3L}(i) - 2q_i \sin^2\theta_W \,, \qquad (10.2a)$$
$$g^i_A \equiv t_{3L}(i) \,, \qquad (10.2b)$$

where $t_{3L}(i)$ is the weak isospin of fermion $i$ ($+1/2$ for $u_i$ and $\nu_i$; $-1/2$ for $d_i$ and $e_i$) and $q_i$ is the charge of $\psi_i$ in units of $e$.

The second term in $\mathscr{L}_F$ represents the charged-current weak interaction [3,4]. For example, the coupling of a $W$ to an electron and a neutrino is

$$-\frac{e}{2\sqrt{2}\sin\theta_W}\left[W_\mu^- \bar{e}\,\gamma^\mu\left(1-\gamma^5\right)\nu + W_\mu^+ \bar{\nu}\,\gamma^\mu\left(1-\gamma^5\right)e\right]. \quad (10.3)$$

For momenta small compared to $M_W$, this term gives rise to the effective four-fermion interaction with the Fermi constant given (at tree level, *i.e.*, lowest order in perturbation theory) by $G_F/\sqrt{2} = g^2/8M_W^2$. $CP$ violation is incorporated in the SM by a single observable phase in $V_{ij}$. The third term in $\mathscr{L}_F$ describes electromagnetic interactions (QED), and the last is the weak neutral-current interaction.

In Eq. (10.1), $m_i$ is the mass of the $i^{th}$ fermion $\psi_i$. For the quarks these are the current masses. For the light quarks, as described in the Particle Listings, $\widehat{m}_u \approx 1.5$–$4.5$ MeV, $\widehat{m}_d \approx 5$–$8.5$ MeV, and $\widehat{m}_s \approx 80$–$155$ MeV. These are running $\overline{\text{MS}}$ masses evaluated at the scale $\mu = 2$ GeV. (In this Section we denote quantities defined in the $\overline{\text{MS}}$ scheme by a caret; the exception is the strong coupling constant, $\alpha_s$, which will always correspond to the $\overline{\text{MS}}$ definition and where the caret will be dropped.) For the heavier quarks we use QCD sum rule constraints [5] and recalculate their masses in each call of our fits to account for their direct $\alpha_s$ dependence. We find, $\widehat{m}_c(\mu = \widehat{m}_c) = 1.290^{+0.040}_{-0.045}$ GeV and $\widehat{m}_b(\mu = \widehat{m}_b) = 4.206 \pm 0.031$ GeV, with a correlation of 29%. The top quark "pole" mass, $m_t = 177.9 \pm 4.4$ GeV, is an average of CDF results from run I [6] and run II [7], as well as the DØ dilepton [8] and lepton plus jets [9] channels. The latter has been recently reanalyzed, leading to a somewhat higher value. We computed the covariance matrix accounting for correlated systematic uncertainties between the different channels and experiments according to Refs. 6 and 10. Our covariance matrix also accounts for a common 0.6 GeV uncertainty (the size of the three-loop term [11]) due to the conversion from the pole mass to the $\overline{\text{MS}}$ mass. We are using a BLM optimized [12] version of the two-loop perturbative QCD formula [13] which should correspond approximately to the kinematic mass extracted from the collider events. The three-loop formula [11] gives virtually identical results. We use $\overline{\text{MS}}$ masses in all expressions to minimize theoretical uncertainties. We will use above value for $m_t$ (together with $M_H = 117$ GeV) for the numerical values quoted in Sec. 10.2–Sec. 10.4. See "The Note on Quark Masses" in the Particle Listings for more information. In the presence of right-handed neutrinos, Eq. (10.1) gives rise also to Dirac neutrino masses. The possibility of Majorana masses is discussed in "Neutrino mass" in the Particle Listings.

$H$ is the physical neutral Higgs scalar which is the only remaining part of $\phi$ after spontaneous symmetry breaking. The Yukawa coupling of $H$ to $\psi_i$, which is flavor diagonal in the minimal model, is $gm_i/2M_W$. In non-minimal models there are additional charged and neutral scalar Higgs particles [14].

## 10.2. Renormalization and radiative corrections

The SM has three parameters (not counting the Higgs boson mass, $M_H$, and the fermion masses and mixings). A particularly useful set is:

(*a*) The fine structure constant $\alpha = 1/137.03599911(46)$, determined from the $e^{\pm}$ anomalous magnetic moment, the quantum Hall effect, and other measurements [15]. In most electroweak renormalization schemes, it is convenient to define a running $\alpha$ dependent on the energy scale of the process, with $\alpha^{-1} \sim 137$ appropriate at very low energy. (The running has also been observed directly [16]. ) For scales above a few hundred MeV this introduces an uncertainty due to the low-energy hadronic contribution to vacuum polarization. In the modified minimal subtraction ($\overline{MS}$) scheme [17] (used for this *Review*), and with $\alpha_s(M_Z) = 0.120$ for the QCD coupling at $M_Z$, we have $\widehat{\alpha}(m_\tau)^{-1} = 133.498 \pm 0.017$ and $\widehat{\alpha}(M_Z)^{-1} = 127.918 \pm 0.018$. These values are updated from Ref. 18 and account for the latest results from $\tau$ decays and a reanalysis of the CMD 2 collaboration results after correcting a radiative correction [19]. See Ref. 20 for a discussion in the context of the anomalous magnetic moment of the muon. The correlation of the latter with $\widehat{\alpha}(M_Z)$, as well as the non-linear $\alpha_s$ dependence of $\widehat{\alpha}(M_Z)$ and the resulting correlation with the input variable $\alpha_s$, are fully taken into account in the fits. The uncertainty is from $e^+e^-$ annihilation data below 1.8 GeV and $\tau$ decay data, from isospin breaking effects (affecting the interpretation of the $\tau$ data), from uncalculated higher order perturbative and non-perturbative QCD corrections, and from the $\overline{MS}$ quark masses. Such a short distance mass definition (unlike the pole mass) is free from non-perturbative and renormalon uncertainties. Various recent evaluations of the contributions of the five light quark flavors, $\Delta\alpha_{\text{had}}^{(5)}$, to the conventional (on-shell) QED coupling, $\alpha(M_Z) = \dfrac{\alpha}{1 - \Delta\alpha}$, are summarized in Table 10.1. Most of the older results relied on $e^+e^- \to$ hadrons cross-section measurements up to energies of 40 GeV, which were somewhat higher than the QCD prediction, suggested stronger running, and were less precise. The most recent results typically assume the validity of perturbative QCD (PQCD) at scales of 1.8 GeV and above, and are in reasonable agreement with each other. (Evaluations in the on-shell scheme utilize resonance data from BES [36] as further input.) There is, however, some discrepancy between analyzes based on $e^+e^- \to$ hadrons cross-section data and those based on $\tau$ decay spectral functions [20]. The latter imply lower central values for the extracted $M_H$ of $\mathcal{O}(10 \text{ GeV})$. Further improvement of this dominant theoretical uncertainty in the interpretation of precision data will require better measurements of the cross-section for $e^+e^- \to$ hadrons below the charmonium resonances, as well as in the threshold region of the heavy quarks (to improve the precision in $\widehat{m}_c(\widehat{m}_c)$ and $\widehat{m}_b(\widehat{m}_b)$). As an alternative to cross-section scans, one can use the high statistics radiative return events [37] at $e^+e^-$ accelerators operating at resonances such as the $\Phi$ or the $\Upsilon(4S)$. The method is systematics dominated. First preliminary results have been presented by the KLOE collaboration [38].

**Table 10.1:** Recent evaluations of the on-shell $\Delta\alpha^{(5)}_{\text{had}}(M_Z)$. For better comparison we adjusted central values and errors to correspond to a common and fixed value of $\alpha_s(M_Z) = 0.120$. References quoting results without the top quark decoupled are converted to the five flavor definition. Ref. [31] uses $\Lambda_{\text{QCD}} = 380 \pm 60$ MeV; for the conversion we assumed $\alpha_s(M_Z) = 0.118 \pm 0.003$.

| Reference | Result | Comment |
|---|---|---|
| Martin & Zeppenfeld [21] | $0.02744 \pm 0.00036$ | PQCD for $\sqrt{s} > 3$ GeV |
| Eidelman & Jegerlehner [22] | $0.02803 \pm 0.00065$ | PQCD for $\sqrt{s} > 40$ GeV |
| Geshkenbein & Morgunov [23] | $0.02780 \pm 0.00006$ | $\mathcal{O}(\alpha_s)$ resonance model |
| Burkhardt & Pietrzyk [24] | $0.0280 \pm 0.0007$ | PQCD for $\sqrt{s} > 40$ GeV |
| Swartz [25] | $0.02754 \pm 0.00046$ | use of fitting function |
| Alemany, Davier, Höcker [26] | $0.02816 \pm 0.00062$ | includes $\tau$ decay data |
| Krasnikov & Rodenberg [27] | $0.02737 \pm 0.00039$ | PQCD for $\sqrt{s} > 2.3$ GeV |
| Davier & Höcker [28] | $0.02784 \pm 0.00022$ | PQCD for $\sqrt{s} > 1.8$ GeV |
| Kühn & Steinhauser [29] | $0.02778 \pm 0.00016$ | complete $\mathcal{O}(\alpha_s^2)$ |
| Erler [18] | $0.02779 \pm 0.00020$ | converted from $\overline{\text{MS}}$ scheme |
| Davier & Höcker [30] | $0.02770 \pm 0.00015$ | use of QCD sum rules |
| Groote *et al.* [31] | $0.02787 \pm 0.00032$ | use of QCD sum rules |
| Martin, Outhwaite, Ryskin [32] | $0.02741 \pm 0.00019$ | includes new BES data |
| Burkhardt & Pietrzyk [33] | $0.02763 \pm 0.00036$ | PQCD for $\sqrt{s} > 12$ GeV |
| de Troconiz & Yndurain [34] | $0.02754 \pm 0.00010$ | PQCD for $s > 2$ GeV$^2$ |
| Jegerlehner [35] | $0.02766 \pm 0.00013$ | converted from MOM scheme |

(b) The Fermi constant, $G_F = 1.16637(1) \times 10^{-5}$ GeV$^{-2}$, determined from the muon lifetime formula [39,40],

$$\tau_\mu^{-1} = \frac{G_F^2 m_\mu^5}{192\pi^3} F\left(\frac{m_e^2}{m_\mu^2}\right)\left(1 + \frac{3}{5}\frac{m_\mu^2}{M_W^2}\right)$$

$$\times \left[1 + \left(\frac{25}{8} - \frac{\pi^2}{2}\right)\frac{\alpha(m_\mu)}{\pi} + C_2\frac{\alpha^2(m_\mu)}{\pi^2}\right], \qquad (10.4a)$$

where

$$F(x) = 1 - 8x + 8x^3 - x^4 - 12x^2 \ln x ,\qquad (10.4b)$$

$$C_2 = \frac{156815}{5184} - \frac{518}{81}\pi^2 - \frac{895}{36}\zeta(3) + \frac{67}{720}\pi^4 + \frac{53}{6}\pi^2 \ln(2) ,\qquad (10.4c)$$

and

$$\alpha(m_\mu)^{-1} = \alpha^{-1} - \frac{2}{3\pi}\ln\left(\frac{m_\mu}{m_e}\right) + \frac{1}{6\pi} \approx 136 . \qquad (10.4d)$$

The $\mathcal{O}(\alpha^2)$ corrections to $\mu$ decay have been completed recently [40]. The remaining uncertainty in $G_F$ is from the experimental input.

(c) The $Z$ boson mass, $M_Z = 91.1876 \pm 0.0021$ GeV, determined from the $Z$-lineshape scan at LEP 1 [41].

With these inputs, $\sin^2\theta_W$ and the $W$ boson mass, $M_W$, can be calculated when values for $m_t$ and $M_H$ are given; conversely (as is done at present), $M_H$ can be constrained by $\sin^2\theta_W$ and $M_W$. The value of $\sin^2\theta_W$ is extracted from $Z$-pole observables and neutral-current processes [41,42], and depends on the renormalization prescription. There are a number of popular schemes [44–50] leading to values which differ by small factors depending on $m_t$ and $M_H$. The notation for these schemes is shown in Table 10.2. Discussion of the schemes follows the table.

Table 10.2: Notations used to indicate the various schemes discussed in the text. Each definition of $\sin\theta_W$ leads to values that differ by small factors depending on $m_t$ and $M_H$.

| Scheme | Notation |
|---|---|
| On-shell | $s_W = \sin\theta_W$ |
| NOV | $s_{M_Z} = \sin\theta_W$ |
| $\overline{\text{MS}}$ | $\widehat{s}_Z = \sin\theta_W$ |
| $\overline{\text{MS}}$ ND | $\widehat{s}_{ND} = \sin\theta_W$ |
| Effective angle | $\overline{s}_f = \sin\theta_W$ |

(i) The on-shell scheme [44] promotes the tree-level formula $\sin^2\theta_W = 1 - M_W^2/M_Z^2$ to a definition of the renormalized $\sin^2\theta_W$ to all orders in perturbation theory, i.e., $\sin^2\theta_W \to s_W^2 \equiv 1 - M_W^2/M_Z^2$:

$$M_W = \frac{A_0}{s_W(1-\Delta r)^{1/2}} , \qquad (10.5a)$$

$$M_Z = \frac{M_W}{c_W} , \qquad (10.5b)$$

where $c_W \equiv \cos\theta_W$, $A_0 = (\pi\alpha/\sqrt{2}G_F)^{1/2} = 37.2805(2)$ GeV, and $\Delta r$ includes the radiative corrections relating $\alpha$, $\alpha(M_Z)$, $G_F$, $M_W$, and $M_Z$. One finds $\Delta r \sim \Delta r_0 - \rho_t/\tan^2\theta_W$, where $\Delta r_0 = 1 - \alpha/\widehat{\alpha}(M_Z) = 0.06654(14)$ is due to the running of $\alpha$, and $\rho_t = 3G_F m_t^2/8\sqrt{2}\pi^2 = 0.00992(m_t/177.9 \text{ GeV})^2$ represents the dominant (quadratic) $m_t$ dependence. There are additional contributions to $\Delta r$ from bosonic loops, including those which depend logarithmically on $M_H$. One has $\Delta r = 0.03434 \mp 0.0017 \pm 0.00014$, where the second uncertainty is from $\alpha(M_Z)$. Thus the value of $s_W^2$ extracted from $M_Z$ includes an uncertainty ($\mp 0.00054$) from the currently allowed range of $m_t$. This scheme is simple conceptually. However, the relatively large ($\sim 3\%$) correction from $\rho_t$ causes large spurious contributions in higher orders.

(ii) A more precisely determined quantity $s_{M_Z}^2$ can be obtained from $M_Z$ by removing the $(m_t, M_H)$ dependent term from $\Delta r$ [45], i.e.,

$$s_{M_Z}^2 c_{M_Z}^2 \equiv \frac{\pi\alpha(M_Z)}{\sqrt{2}\,G_F\,M_Z^2} \ . \tag{10.6}$$

Using $\alpha(M_Z)^{-1} = 128.91 \pm 0.02$ yields $s_{M_Z}^2 = 0.23108 \mp 0.00005$. The small uncertainty in $s_{M_Z}^2$ compared to other schemes is because most of the $m_t$ dependence has been removed by definition. However, the $m_t$ uncertainty reemerges when other quantities (e.g., $M_W$ or other $Z$-pole observables) are predicted in terms of $M_Z$.

Both $s_W^2$ and $s_{M_Z}^2$ depend not only on the gauge couplings but also on the spontaneous-symmetry breaking, and both definitions are awkward in the presence of any extension of the SM which perturbs the value of $M_Z$ (or $M_W$). Other definitions are motivated by the tree-level coupling constant definition $\theta_W = \tan^{-1}(g'/g)$.

(iii) In particular, the modified minimal subtraction ($\overline{\text{MS}}$) scheme introduces the quantity $\sin^2\widehat{\theta}_W(\mu) \equiv \widehat{g}'^2(\mu)/[\widehat{g}^2(\mu) + \widehat{g}'^2(\mu)]$, where the couplings $\widehat{g}$ and $\widehat{g}'$ are defined by modified minimal subtraction and the scale $\mu$ is conveniently chosen to be $M_Z$ for many electroweak processes. The value of $\widehat{s}_Z^2 = \sin^2\widehat{\theta}_W(M_Z)$ extracted from $M_Z$ is less sensitive than $s_W^2$ to $m_t$ (by a factor of $\tan^2\theta_W$), and is less sensitive to most types of new physics than $s_W^2$ or $s_{M_Z}^2$. It is also very useful for comparing with the predictions of grand unification. There are actually several variant definitions of $\sin^2\widehat{\theta}_W(M_Z)$, differing according to whether or how finite $\alpha\ln(m_t/M_Z)$ terms are decoupled (subtracted from the couplings). One cannot entirely decouple the $\alpha\ln(m_t/M_Z)$ terms from all electroweak quantities because $m_t \gg m_b$ breaks SU(2) symmetry. The scheme that will be adopted here decouples the $\alpha\ln(m_t/M_Z)$ terms from the $\gamma$–$Z$ mixing [17,46], essentially eliminating any $\ln(m_t/M_Z)$ dependence in the formulae for asymmetries at the $Z$-pole when written in terms of $\widehat{s}_Z^2$. (A similar definition is used for $\widehat{\alpha}$.) The various definitions are related by

$$\widehat{s}_Z^2 = c\,(m_t, M_H)\,s_W^2 = \overline{c}\,(m_t, M_H)\,s_{M_Z}^2\,, \tag{10.7}$$

where $c = 1.0381 \pm 0.0019$ and $\overline{c} = 1.0003 \mp 0.0006$. The quadratic $m_t$ dependence is given by $c \sim 1 + \rho_t/\tan^2\theta_W$ and $\overline{c} \sim 1 - \rho_t/(1 - \tan^2\theta_W)$, respectively. The expressions for $M_W$ and $M_Z$ in the $\overline{MS}$ scheme are

$$M_W = \frac{A_0}{\widehat{s}_Z\,(1 - \Delta\widehat{r}_W)^{1/2}}\,, \tag{10.8a}$$

$$M_Z = \frac{M_W}{\widehat{\rho}^{1/2}\widehat{c}_Z}\,, \tag{10.8b}$$

and one predicts $\Delta\widehat{r}_W = 0.06976 \pm 0.00006 \pm 0.00014$. $\Delta\widehat{r}_W$ has no quadratic $m_t$ dependence, because shifts in $M_W$ are absorbed into the observed $G_F$, so that the error in $\Delta\widehat{r}_W$ is dominated by $\Delta r_0 = 1 - \alpha/\widehat{\alpha}(M_Z)$ which induces the second quoted uncertainty. The quadratic $m_t$ dependence has been shifted into $\widehat{\rho} \sim 1 + \rho_t$, where including bosonic loops, $\widehat{\rho} = 1.0110 \pm 0.0005$.

(iv) A variant $\overline{MS}$ quantity $\widehat{s}_{ND}^2$ (used in the 1992 edition of this *Review*) does not decouple the $\alpha\ln(m_t/M_Z)$ terms [47]. It is related to $\widehat{s}_Z^2$ by

$$\widehat{s}_Z^2 = \widehat{s}_{ND}^2 / \left(1 + \frac{\widehat{\alpha}}{\pi}d\right)\,, \tag{10.9a}$$

$$d = \frac{1}{3}\left(\frac{1}{\widehat{s}^2} - \frac{8}{3}\right)\left[\left(1 + \frac{\alpha_s}{\pi}\right)\ln\frac{m_t}{M_Z} - \frac{15\alpha_s}{8\pi}\right]\,, \tag{10.9b}$$

Thus, $\widehat{s}_Z^2 - \widehat{s}_{ND}^2 \sim -0.0002$ for $m_t = 177.9$ GeV.

(v) Yet another definition, the effective angle [48–50] $\overline{s}_f^2$ for the $Z$ vector coupling to fermion $f$, is described in Sec. 10.3.

Experiments are at a level of precision that complete $\mathcal{O}(\alpha)$ radiative corrections must be applied. For neutral-current and $Z$-pole processes, these corrections are conveniently divided into two classes:

1. QED diagrams involving the emission of real photons or the exchange of virtual photons in loops, but not including vacuum polarization diagrams. These graphs often yield finite and gauge-invariant contributions to observable processes. However, they are dependent on energies, experimental cuts, *etc.*, and must be calculated individually for each experiment.

2. Electroweak corrections, including $\gamma\gamma$, $\gamma Z$, $ZZ$, and $WW$ vacuum polarization diagrams, as well as vertex corrections, box graphs, *etc.*, involving virtual $W$'s and $Z$'s. Many of these corrections are absorbed into the renormalized Fermi constant defined in Eq. (10.4). Others modify the tree-level expressions for $Z$-pole observables and neutral-current amplitudes in several ways [42].

One-loop corrections are included for all processes. In addition, certain two-loop corrections are also important. In particular, two-loop corrections involving the top quark modify $\rho_t$ in $\hat{\rho}$, $\Delta r$, and elsewhere by

$$\rho_t \to \rho_t \left[1 + R\left(M_H, m_t\right) \rho_t / 3\right] . \tag{10.10}$$

$R(M_H, m_t)$ is best described as an expansion in $M_Z^2/m_t^2$. The unsuppressed terms were first obtained in Ref. 51, and are known analytically [52]. Contributions suppressed by $M_Z^2/m_t^2$ were first studied in Ref. 53 with the help of small and large Higgs mass expansions, which can be interpolated. These contributions are about as large as the leading ones in Refs. 51 and 52. In addition, the complete two-loop calculation of diagrams containing at least one fermion loop and contributing to $\Delta r$ has been performed without further approximation in Ref. 54. The two-loop evaluation of $\Delta r$ was completed with the purely bosonic contributions in Ref. 55. For $M_H$ above its lower direct limit, $-17 < R \leq -13$. Mixed QCD-electroweak loops of order $\alpha\alpha_s m_t^2$ [56] and $\alpha\alpha_s^2 m_t^2$ [57] increase the predicted value of $m_t$ by 6%. This is, however, almost entirely an artifact of using the pole mass definition for $m_t$. The equivalent corrections when using the $\overline{\rm MS}$ definition $\widehat{m}_t(\widehat{m}_t)$ increase $m_t$ by less than 0.5%. The leading electroweak [51,52] and mixed [58] two-loop terms are also known for the $Z \to b\bar{b}$ vertex, but not the respective subleading ones. $\mathcal{O}(\alpha\alpha_s)$-vertex corrections involving massless quarks have been obtained in Ref. [59]. Since they add coherently, the resulting effect is sizable, and shifts the extracted $\alpha_s(M_Z)$ by $\approx +0.0007$. Corrections of the same order to $Z \to b\bar{b}$ decays have also been completed [60].

Throughout this *Review* we utilize electroweak radiative corrections from the program GAPP [61], which works entirely in the $\overline{\rm MS}$ scheme, and which is independent of the package ZFITTER [50].

### 10.3. Cross-section and asymmetry formulas

It is convenient to write the four-fermion interactions relevant to $\nu$-hadron, $\nu$-$e$, and parity violating $e$-hadron neutral-current processes in a form that is valid in an arbitrary gauge theory (assuming massless left-handed neutrinos). One has

$$-\mathcal{L}^{\nu \text{Hadron}} = \frac{G_F}{\sqrt{2}} \bar{\nu} \gamma^\mu \left(1 - \gamma^5\right) \nu$$

$$\times \sum_i \left[\epsilon_L(i) \bar{q}_i \gamma_\mu \left(1 - \gamma^5\right) q_i + \epsilon_R(i) \bar{q}_i \gamma_\mu \left(1 + \gamma^5\right) q_i\right] , \tag{10.11}$$

$$-\mathcal{L}^{\nu e} = \frac{G_F}{\sqrt{2}} \bar{\nu}_\mu \gamma^\mu \left(1 - \gamma^5\right) \nu_\mu \, \bar{e} \, \gamma_\mu \left(g_V^{\nu e} - g_A^{\nu e} \gamma^5\right) e \tag{10.12}$$

(for $\nu_e$-$e$ or $\bar\nu_e$-$e$, the charged-current contribution must be included), and

$$-\mathscr{L}^{e\text{Hadron}} = -\frac{G_F}{\sqrt{2}}$$

$$\times \sum_i \left[ C_{1i}\, \bar{e}\, \gamma_\mu\, \gamma^5\, e\, \bar{q}_i\, \gamma^\mu\, q_i + C_{2i}\, \bar{e}\, \gamma_\mu\, e\, \bar{q}_i\, \gamma^\mu\, \gamma^5\, q_i \right]. \quad (10.13)$$

(One must add the parity-conserving QED contribution.)

The SM expressions for $\epsilon_{L,R}(i)$, $g^{\nu e}_{V,A}$, and $C_{ij}$ are given in Table 10.3. Note, that $g^{\nu e}_{V,A}$ and the other quantities are coefficients of effective four-Fermi operators, which differ from the quantities defined in Eq. (10.2) in the radiative corrections and in the presence of possible physics beyond the SM.

A precise determination of the on-shell $s_W^2$, which depends only very weakly on $m_t$ and $M_H$, is obtained from deep inelastic neutrino scattering from (approximately) isoscalar targets [62]. The ratio $R_\nu \equiv \sigma^{NC}_{\nu N}/\sigma^{CC}_{\nu N}$ of neutral- to charged-current cross-sections has been measured to 1% accuracy by the CDHS [63] and CHARM [64] collaborations at CERN, and the CCFR [65] collaboration at Fermilab has obtained an even more precise result, so it is important to obtain theoretical expressions for $R_\nu$ and $R_{\bar\nu} \equiv \sigma^{NC}_{\bar\nu N}/\sigma^{CC}_{\bar\nu N}$ to comparable accuracy. Fortunately, most of the uncertainties from the strong interactions and neutrino spectra cancel in the ratio. The largest theoretical uncertainty is associated with the $c$-threshold, which mainly affects $\sigma^{CC}$. Using the slow rescaling prescription [66] the central value of $\sin^2\theta_W$ from CCFR varies as $0.0111(m_c\,[\text{GeV}] - 1.31)$, where $m_c$ is the effective mass which is numerically close to the $\overline{\text{MS}}$ mass $\widehat{m}_c(\widehat{m}_c)$, but their exact relation is unknown at higher orders. For $m_c = 1.31 \pm 0.24$ GeV (determined from $\nu$-induced dimuon production [67]) this contributes $\pm 0.003$ to the total uncertainty $\Delta \sin^2\theta_W \sim \pm 0.004$. (The experimental uncertainty is also $\pm 0.003$.) This uncertainty largely cancels, however, in the Paschos-Wolfenstein ratio [68],

$$R^- = \frac{\sigma^{NC}_{\nu N} - \sigma^{NC}_{\bar\nu N}}{\sigma^{CC}_{\nu N} - \sigma^{CC}_{\bar\nu N}}. \quad (10.14)$$

It was measured recently by the NuTeV collaboration [69] for the first time, and required a high-intensity and high-energy anti-neutrino beam.

A simple zero$^{th}$-order approximation is

$$R_\nu = g_L^2 + g_R^2 r, \quad (10.15a)$$

$$R_{\bar\nu} = g_L^2 + \frac{g_R^2}{r}, \quad (10.15b)$$

$$R^- = g_L^2 - g_R^2, \quad (10.15c)$$

where

$$g_L^2 \equiv \epsilon_L(u)^2 + \epsilon_L(d)^2 \approx \frac{1}{2} - \sin^2\theta_W + \frac{5}{9}\sin^4\theta_W, \quad (10.16a)$$

**Table 10.3:** Standard Model expressions for the neutral-current parameters for $\nu$-hadron, $\nu$-$e$, and $e$-hadron processes. At tree level, $\rho = \kappa = 1$, $\lambda = 0$. If radiative corrections are included, $\rho_{\nu N}^{NC} = 1.0086$, $\widehat{\kappa}_{\nu N}(\langle Q^2 \rangle = -12 \text{ GeV}^2) = 0.9978$, $\widehat{\kappa}_{\nu N}(\langle Q^2 \rangle = -35 \text{ GeV}^2) = 0.9965$, $\lambda_{uL} = -0.0031$, $\lambda_{dL} = -0.0025$, and $\lambda_{dR} = 2\lambda_{uR} = 7.5 \times 10^{-5}$. For $\nu$-$e$ scattering, $\rho_{\nu e} = 1.0132$ and $\widehat{\kappa}_{\nu e} = 0.9967$ (at $\langle Q^2 \rangle = 0$.). For atomic parity violation and the SLAC polarized electron experiment, $\rho'_{eq} = 0.9881$, $\rho_{eq} = 1.0011$, $\widehat{\kappa}'_{eq} = 1.0027$, $\widehat{\kappa}_{eq} = 1.0300$, $\lambda_{1d} = -2\lambda_{1u} = 3.7 \times 10^{-5}$, $\lambda_{2u} = -0.0121$ and $\lambda_{2d} = 0.0026$. The dominant $m_t$ dependence is given by $\rho \sim 1 + \rho_t$, while $\widehat{\kappa} \sim 1$ ($\overline{\text{MS}}$) or $\kappa \sim 1 + \rho_t/\tan^2\theta_W$ (on-shell).

| Quantity | Standard Model Expression |
|---|---|
| $\epsilon_L(u)$ | $\rho_{\nu N}^{NC} \left( \frac{1}{2} - \frac{2}{3}\widehat{\kappa}_{\nu N}\,\widehat{s}_Z^2 \right) + \lambda_{uL}$ |
| $\epsilon_L(d)$ | $\rho_{\nu N}^{NC} \left( -\frac{1}{2} + \frac{1}{3}\widehat{\kappa}_{\nu N}\,\widehat{s}_Z^2 \right) + \lambda_{dL}$ |
| $\epsilon_R(u)$ | $\rho_{\nu N}^{NC} \left( -\frac{2}{3}\widehat{\kappa}_{\nu N}\,\widehat{s}_Z^2 \right) + \lambda_{uR}$ |
| $\epsilon_R(d)$ | $\rho_{\nu N}^{NC} \left( \frac{1}{3}\widehat{\kappa}_{\nu N}\,\widehat{s}_Z^2 \right) + \lambda_{dR}$ |
| $g_V^{\nu e}$ | $\rho_{\nu e} \left( -\frac{1}{2} + 2\widehat{\kappa}_{\nu e}\,\widehat{s}_Z^2 \right)$ |
| $g_A^{\nu e}$ | $\rho_{\nu e} \left( -\frac{1}{2} \right)$ |
| $C_{1u}$ | $\rho'_{eq} \left( -\frac{1}{2} + \frac{4}{3}\widehat{\kappa}'_{eq}\,\widehat{s}_Z^2 \right) + \lambda_{1u}$ |
| $C_{1d}$ | $\rho'_{eq} \left( \frac{1}{2} - \frac{2}{3}\widehat{\kappa}'_{eq}\,\widehat{s}_Z^2 \right) + \lambda_{1d}$ |
| $C_{2u}$ | $\rho_{eq} \left( -\frac{1}{2} + 2\widehat{\kappa}_{eq}\,\widehat{s}_Z^2 \right) + \lambda_{2u}$ |
| $C_{2d}$ | $\rho_{eq} \left( \frac{1}{2} - 2\widehat{\kappa}_{eq}\,\widehat{s}_Z^2 \right) + \lambda_{2d}$ |

$$g_R^2 \equiv \epsilon_R(u)^2 + \epsilon_R(d)^2 \approx \frac{5}{9}\sin^4\theta_W \,, \qquad (10.16b)$$

and $r \equiv \sigma_{\bar{\nu}N}^{CC}/\sigma_{\nu N}^{CC}$ is the ratio of $\bar{\nu}$ and $\nu$ charged-current cross-sections, which can be measured directly. (In the simple parton model, ignoring hadron energy cuts, $r \approx (\frac{1}{3} + \epsilon)/(1 + \frac{1}{3}\epsilon)$, where $\epsilon \sim 0.125$ is the ratio of the fraction of the nucleon's momentum carried by antiquarks to that carried by quarks.) In practice, Eq. (10.15) must be corrected for quark mixing, quark sea effects, $c$-quark threshold effects, non-isoscalarity, $W$–$Z$ propagator differences, the finite muon mass, QED and electroweak radiative corrections. Details of the neutrino spectra, experimental cuts, $x$ and $Q^2$ dependence of structure functions, and longitudinal structure functions enter only at the level of these corrections and therefore lead to very small uncertainties. The CCFR group quotes $s_W^2 = 0.2236 \pm 0.0041$ for $(m_t, M_H) = (175, 150)$ GeV with very little sensitivity to $(m_t, M_H)$. The NuTeV collaboration finds $s_W^2 = 0.2277 \pm 0.0016$ (for the same reference values) which is 3.0 $\sigma$ higher than the SM prediction. The discrepancy is in the left-handed coupling, $g_L^2 = 0.3000 \pm 0.0014$, which is 2.9 $\sigma$ low, while $g_R^2 = 0.0308 \pm 0.0011$ is 0.6 $\sigma$ high. It is conceivable that the effect is caused by an asymmetric strange sea [70]. A preliminary analysis of dimuon data [71] in the relevant kinematic regime, however, indicates an asymmetric strange sea with the wrong sign to explain the discrepancy [72]. Another possibility is that the parton distribution functions (PDFs) violate isospin symmetry at levels much stronger than generally expected. Isospin breaking, nuclear physics, and higher order QCD effects seem unlikely explanations of the NuTeV discrepancy but need further study. The extracted $g_{L,R}^2$ may also shift if analyzed using the most recent set of QED and electroweak radiative corrections [73].

The laboratory cross-section for $\nu_\mu e \to \nu_\mu e$ or $\overline{\nu}_\mu e \to \overline{\nu}_\mu e$ elastic scattering is

$$\frac{d\sigma_{\nu_\mu, \overline{\nu}_\mu}}{dy} = \frac{G_F^2 m_e E_\nu}{2\pi}$$

$$\times \left[ (g_V^{\nu e} \pm g_A^{\nu e})^2 + (g_V^{\nu e} \mp g_A^{\nu e})^2 (1-y)^2 \right.$$

$$\left. - \left(g_V^{\nu e 2} - g_A^{\nu e 2}\right) \frac{y\, m_e}{E_\nu} \right], \tag{10.17}$$

where the upper (lower) sign refers to $\nu_\mu (\overline{\nu}_\mu)$, and $y \equiv E_e/E_\nu$ (which runs from 0 to $(1 + m_e/2E_\nu)^{-1}$) is the ratio of the kinetic energy of the recoil electron to the incident $\nu$ or $\overline{\nu}$ energy. For $E_\nu \gg m_e$ this yields a total cross-section

$$\sigma = \frac{G_F^2\, m_e\, E_\nu}{2\pi} \left[ (g_V^{\nu e} \pm g_A^{\nu e})^2 + \frac{1}{3} (g_V^{\nu e} \mp g_A^{\nu e})^2 \right]. \tag{10.18}$$

The most accurate leptonic measurements [74–77] of $\sin^2\theta_W$ are from the ratio $R \equiv \sigma_{\nu_\mu e}/\sigma_{\overline{\nu}_\mu e}$ in which many of the systematic uncertainties cancel. Radiative corrections (other than $m_t$ effects) are small compared to the precision of present experiments and have negligible effect on the extracted $\sin^2\theta_W$. The most precise experiment (CHARM II) [76] determined not only $\sin^2\theta_W$ but $g_{V,A}^{\nu e}$ as well. The cross-sections for $\nu_e$-$e$ and $\overline{\nu}_e$-$e$ may be obtained from Eq. (10.17) by replacing $g_{V,A}^{\nu e}$ by $g_{V,A}^{\nu e} + 1$, where the 1 is due to the charged-current contribution [77,78].

The SLAC polarized-electron experiment [79] measured the parity-violating asymmetry

$$A = \frac{\sigma_R - \sigma_L}{\sigma_R + \sigma_L}, \tag{10.19}$$

where $\sigma_{R,L}$ is the cross-section for the deep-inelastic scattering of a right- or left-handed electron: $e_{R,L}N \to eX$. In the quark parton model

$$\frac{A}{Q^2} = a_1 + a_2 \frac{1-(1-y)^2}{1+(1-y)^2} \;, \quad (10.20)$$

where $Q^2 > 0$ is the momentum transfer and $y$ is the fractional energy transfer from the electron to the hadrons. For the deuteron or other isoscalar targets, one has, neglecting the $s$-quark and antiquarks,

$$a_1 = \frac{3G_F}{5\sqrt{2}\pi\alpha}\left(C_{1u} - \frac{1}{2}C_{1d}\right) \approx \frac{3G_F}{5\sqrt{2}\pi\alpha}\left(-\frac{3}{4} + \frac{5}{3}\sin^2\theta_W\right) \;, \quad (10.21a)$$

$$a_2 = \frac{3G_F}{5\sqrt{2}\pi\alpha}\left(C_{2u} - \frac{1}{2}C_{2d}\right) \approx \frac{9G_F}{5\sqrt{2}\pi\alpha}\left(\sin^2\theta_W - \frac{1}{4}\right) \;. \quad (10.21b)$$

There are now precise experiments measuring atomic parity violation [80] in cesium (at the 0.4% level) [81], thallium [82], lead [83], and bismuth [84]. The uncertainties associated with atomic wave functions are quite small for cesium [85], and have been reduced recently to about 0.4% [86]. In the past, the semi-empirical value of the tensor polarizability added another source of theoretical uncertainty [87]. The ratio of the off-diagonal hyperfine amplitude to the polarizability has now been measured directly by the Boulder group [86]. Combined with the precisely known hyperfine amplitude [88] one finds excellent agreement with the earlier results, reducing the overall theory uncertainty to only 0.5% (while slightly increasing the experimental error). An earlier 2.3 $\sigma$ deviation from the SM (see the year 2000 edition of this *Review*) is now seen at the 1 $\sigma$ level, after the contributions from the Breit interaction have been reevaluated [89], and after the subsequent inclusion of other large and previously underestimated effects [90] (*e.g.*, from QED radiative corrections), and an update of the SM calculation [91] resulted in a vanishing net effect. The theoretical uncertainties are 3% for thallium [92] but larger for the other atoms. For heavy atoms one determines the "weak charge"

$$Q_W = -2\left[C_{1u}\left(2Z+N\right) + C_{1d}\left(Z+2N\right)\right]$$
$$\approx Z\left(1 - 4\sin^2\theta_W\right) - N \;. \quad (10.22)$$

The recent Boulder experiment in cesium also observed the parity-violating weak corrections to the nuclear electromagnetic vertex (the anapole moment [93]).

In the future it could be possible to reduce the theoretical wave function uncertainties by taking the ratios of parity violation in different isotopes [80,94]. There would still be some residual uncertainties from differences in the neutron charge radii, however [95].

The forward-backward asymmetry for $e^+e^- \to \ell^+\ell^-$, $\ell = \mu$ or $\tau$, is defined as

$$A_{FB} \equiv \frac{\sigma_F - \sigma_B}{\sigma_F + \sigma_B} \;, \quad (10.23)$$

*10. EW model and constraints on new physics* 13

where $\sigma_F(\sigma_B)$ is the cross-section for $\ell^-$ to travel forward (backward) with respect to the $e^-$ direction. $A_{FB}$ and $R$, the total cross-section relative to pure QED, are given by

$$R = F_1 \,, \quad (10.24)$$

$$A_{FB} = 3F_2/4F_1 \,, \quad (10.25)$$

where

$$F_1 = 1 - 2\chi_0\, g_V^e\, g_V^\ell\, \cos\delta_R + \chi_0^2 \left(g_V^{e2} + g_A^{e2}\right)\left(g_V^{\ell 2} + g_A^{\ell 2}\right), \quad (10.26a)$$

$$F_2 = -2\chi_0\, g_A^e\, g_A^\ell\, \cos\delta_R + 4\chi_0^2\, g_A^e\, g_A^\ell\, g_V^e\, g_V^\ell \,, \quad (10.26b)$$

$$\tan\delta_R = \frac{M_Z \Gamma_Z}{M_Z^2 - s} \,, \quad (10.27)$$

$$\chi_0 = \frac{G_F}{2\sqrt{2}\pi\alpha} \frac{sM_Z^2}{\left[(M_Z^2 - s)^2 + M_Z^2\Gamma_Z^2\right]^{1/2}} \,, \quad (10.28)$$

and $\sqrt{s}$ is the CM energy. Eq. (10.26) is valid at tree level. If the data is radiatively corrected for QED effects (as described above), then the remaining electroweak corrections can be incorporated [96,97] (in an approximation adequate for existing PEP, PETRA, and TRISTAN data, which are well below the $Z$-pole) by replacing $\chi_0$ by $\chi(s) \equiv (1 + \rho_t)\chi_0(s)\alpha/\alpha(s)$, where $\alpha(s)$ is the running QED coupling, and evaluating $g_V$ in the $\overline{\text{MS}}$ scheme. Formulas for $e^+e^- \to$ hadrons may be found in Ref. 98.

At LEP and SLC, there were high-precision measurements of various $Z$-pole observables [41,99–105], as summarized in Table 10.4. These include the $Z$ mass and total width, $\Gamma_Z$, and partial widths $\Gamma(f\bar{f})$ for $Z \to f\bar{f}$ where fermion $f = e$, $\mu$, $\tau$, hadrons, $b$, or $c$. It is convenient to use the variables $M_Z$, $\Gamma_Z$, $R_\ell \equiv \Gamma(\text{had})/\Gamma(\ell^+\ell^-)$, $\sigma_{\text{had}} \equiv 12\pi\Gamma(e^+e^-)\Gamma(\text{had})/M_Z^2\Gamma_Z^2$, $R_b \equiv \Gamma(b\bar{b})/\Gamma(\text{had})$, and $R_c \equiv \Gamma(c\bar{c})/\Gamma(\text{had})$, most of which are weakly correlated experimentally. ($\Gamma(\text{had})$ is the partial width into hadrons.) $\mathcal{O}(\alpha^3)$ QED corrections introduce a large anticorrelation ($-30\%$) between $\Gamma_Z$ and $\sigma_{\text{had}}$ [41], while the anticorrelation between $R_b$ and $R_c$ ($-14\%$) is smaller than previously [100]. $R_\ell$ is insensitive to $m_t$ except for the $Z \to b\bar{b}$ vertex and final state corrections and the implicit dependence through $\sin^2\theta_W$. Thus it is especially useful for constraining $\alpha_s$. The width for invisible decays [41], $\Gamma(\text{inv}) = \Gamma_Z - 3\Gamma(\ell^+\ell^-) - \Gamma(\text{had}) = 499.0 \pm 1.5$ MeV, can be used to determine the number of neutrino flavors much lighter than $M_Z/2$, $N_\nu = \Gamma(\text{inv})/\Gamma^{\text{theory}}(\nu\bar{\nu}) = 2.983 \pm 0.009$ for $(m_t, M_H) = (177.9, 117)$ GeV.

There were also measurements of various $Z$-pole asymmetries. These include the polarization or left-right asymmetry

$$A_{LR} \equiv \frac{\sigma_L - \sigma_R}{\sigma_L + \sigma_R} \,, \quad (10.29)$$

where $\sigma_F(\sigma_B)$ is the cross-section for $\ell^-$ to travel forward (backward) with respect to the $e^-$ direction. $A_{FB}$ and $R$, the total cross-section relative to pure QED, are given by

$$R = F_1 \,, \quad (10.24)$$

$$A_{FB} = 3F_2/4F_1 \,, \quad (10.25)$$

where

$$F_1 = 1 - 2\chi_0\, g_V^e\, g_V^\ell\, \cos\delta_R + \chi_0^2 \left(g_V^{e2} + g_A^{e2}\right)\left(g_V^{\ell 2} + g_A^{\ell 2}\right), \quad (10.26a)$$

$$F_2 = -2\chi_0\, g_A^e\, g_A^\ell\, \cos\delta_R + 4\chi_0^2\, g_A^e\, g_A^\ell\, g_V^e\, g_V^\ell \,, \quad (10.26b)$$

$$\tan\delta_R = \frac{M_Z \Gamma_Z}{M_Z^2 - s} \,, \quad (10.27)$$

$$\chi_0 = \frac{G_F}{2\sqrt{2}\pi\alpha} \frac{sM_Z^2}{\left[(M_Z^2 - s)^2 + M_Z^2\Gamma_Z^2\right]^{1/2}} \,, \quad (10.28)$$

and $\sqrt{s}$ is the CM energy. Eq. (10.26) is valid at tree level. If the data is radiatively corrected for QED effects (as described above), then the remaining electroweak corrections can be incorporated [96,97] (in an approximation adequate for existing PEP, PETRA, and TRISTAN data, which are well below the $Z$-pole) by replacing $\chi_0$ by $\chi(s) \equiv (1 + \rho_t)\chi_0(s)\alpha/\alpha(s)$, where $\alpha(s)$ is the running QED coupling, and evaluating $g_V$ in the $\overline{\text{MS}}$ scheme. Formulas for $e^+e^- \to$ hadrons may be found in Ref. 98.

At LEP and SLC, there were high-precision measurements of various $Z$-pole observables [41,99–105], as summarized in Table 10.4. These include the $Z$ mass and total width, $\Gamma_Z$, and partial widths $\Gamma(f\bar{f})$ for $Z \to f\bar{f}$ where fermion $f = e$, $\mu$, $\tau$, hadrons, $b$, or $c$. It is convenient to use the variables $M_Z$, $\Gamma_Z$, $R_\ell \equiv \Gamma(\text{had})/\Gamma(\ell^+\ell^-)$, $\sigma_{\text{had}} \equiv 12\pi\Gamma(e^+e^-)\Gamma(\text{had})/M_Z^2\Gamma_Z^2$, $R_b \equiv \Gamma(b\bar{b})/\Gamma(\text{had})$, and $R_c \equiv \Gamma(c\bar{c})/\Gamma(\text{had})$, most of which are weakly correlated experimentally. ($\Gamma(\text{had})$ is the partial width into hadrons.) $\mathcal{O}(\alpha^3)$ QED corrections introduce a large anticorrelation ($-30\%$) between $\Gamma_Z$ and $\sigma_{\text{had}}$ [41], while the anticorrelation between $R_b$ and $R_c$ ($-14\%$) is smaller than previously [100]. $R_\ell$ is insensitive to $m_t$ except for the $Z \to b\bar{b}$ vertex and final state corrections and the implicit dependence through $\sin^2\theta_W$. Thus it is especially useful for constraining $\alpha_s$. The width for invisible decays [41], $\Gamma(\text{inv}) = \Gamma_Z - 3\Gamma(\ell^+\ell^-) - \Gamma(\text{had}) = 499.0 \pm 1.5$ MeV, can be used to determine the number of neutrino flavors much lighter than $M_Z/2$, $N_\nu = \Gamma(\text{inv})/\Gamma^{\text{theory}}(\nu\bar{\nu}) = 2.983 \pm 0.009$ for $(m_t, M_H) = (177.9, 117)$ GeV.

There were also measurements of various $Z$-pole asymmetries. These include the polarization or left-right asymmetry

$$A_{LR} \equiv \frac{\sigma_L - \sigma_R}{\sigma_L + \sigma_R} \,, \quad (10.29)$$

where $\sigma_L(\sigma_R)$ is the cross-section for a left-(right-)handed incident electron. $A_{LR}$ has been measured precisely by the SLD collaboration at the SLC [101], and has the advantages of being extremely sensitive to $\sin^2\theta_W$ and that systematic uncertainties largely cancel. In addition, the SLD collaboration has extracted the final-state couplings $A_b$, $A_c$ [41], $A_s$ [102], $A_\tau$, and $A_\mu$ [103] from left-right forward-backward asymmetries, using

$$A_{LR}^{FB}(f) = \frac{\sigma_{LF}^f - \sigma_{LB}^f - \sigma_{RF}^f + \sigma_{RB}^f}{\sigma_{LF}^f + \sigma_{LB}^f + \sigma_{RF}^f + \sigma_{RB}^f} = \frac{3}{4}A_f , \qquad (10.30)$$

where, for example, $\sigma_{LF}$ is the cross-section for a left-handed incident electron to produce a fermion $f$ traveling in the forward hemisphere. Similarly, $A_\tau$ is measured at LEP [41] through the negative total $\tau$ polarization, $\mathcal{P}_\tau$, and $A_e$ is extracted from the angular distribution of $\mathcal{P}_\tau$. An equation such as (10.30) assumes that initial state QED corrections, photon exchange, $\gamma$–$Z$ interference, the tiny electroweak boxes, and corrections for $\sqrt{s} \neq M_Z$ are removed from the data, leaving the pure electroweak asymmetries. This allows the use of effective tree-level expressions,

$$A_{LR} = A_e P_e , \qquad (10.31)$$

$$A_{FB} = \frac{3}{4}A_f \frac{A_e + P_e}{1 + P_e A_e} , \qquad (10.32)$$

where

$$A_f \equiv \frac{2\overline{g}_V^f \, \overline{g}_A^f}{\overline{g}_V^{f2} + \overline{g}_A^{f2}} , \qquad (10.33)$$

and

$$\overline{g}_V^f = \sqrt{\rho_f} \left( t_{3L}^{(f)} - 2q_f \kappa_f \sin^2\theta_W \right) , \qquad (10.33b)$$

$$\overline{g}_A^f = \sqrt{\rho_f} \, t_{3L}^{(f)} . \qquad (10.33c)$$

$P_e$ is the initial $e^-$ polarization, so that the second equality in Eq. (10.30) is reproduced for $P_e = 1$, and the $Z$-pole forward-backward asymmetries at LEP ($P_e = 0$) are given by $A_{FB}^{(0,f)} = \frac{3}{4}A_e A_f$ where $f = e$, $\mu$, $\tau$, $b$, $c$, $s$ [104], and $q$, and where $A_{FB}^{(0,q)}$ refers to the hadronic charge asymmetry. Corrections for $t$-channel exchange and $s/t$-channel interference cause $A_{FB}^{(0,e)}$ to be strongly anticorrelated with $R_e$ ($-37\%$). The initial state coupling, $A_e$, is also determined through the left-right charge asymmetry [105] and in polarized Bhabba scattering at the SLC [103].

The electroweak radiative corrections have been absorbed into corrections $\rho_f - 1$ and $\kappa_f - 1$, which depend on the fermion $f$ and on the renormalization scheme. In the on-shell scheme, the quadratic $m_t$ dependence is given by $\rho_f \sim 1 + \rho_t$, $\kappa_f \sim 1 + \rho_t/\tan^2\theta_W$, while in $\overline{\text{MS}}$, $\widehat{\rho}_f \sim \widehat{\kappa}_f \sim 1$, for $f \neq b$ ($\widehat{\rho}_b \sim 1 - \frac{4}{3}\rho_t$, $\widehat{\kappa}_b \sim 1 + \frac{2}{3}\rho_t$). In the $\overline{\text{MS}}$ scheme

*10. EW model and constraints on new physics* **15**

the normalization is changed according to $G_F M_Z^2/2\sqrt{2}\pi \to \widehat{\alpha}/4\widehat{s}_Z^2\widehat{c}_Z^2$. (If one continues to normalize amplitudes by $G_F M_Z^2/2\sqrt{2}\pi$, as in the 1996 edition of this *Review*, then $\widehat{\rho}_f$ contains an additional factor of $\widehat{\rho}$.) In practice, additional bosonic and fermionic loops, vertex corrections, leading higher order contributions, *etc.*, must be included. For example, in the $\overline{\text{MS}}$ scheme one has $\widehat{\rho}_\ell = 0.9981$, $\widehat{\kappa}_\ell = 1.0013$, $\widehat{\rho}_b = 0.9861$, and $\widehat{\kappa}_b = 1.0071$. It is convenient to define an effective angle $\overline{s}_f^2 \equiv \sin^2 \overline{\theta}_{Wf} \equiv \widehat{\kappa}_f \widehat{s}_Z^2 = \kappa_f s_W^2$, in terms of which $\overline{g}_V^f$ and $\overline{g}_A^f$ are given by $\sqrt{\rho_f}$ times their tree-level formulae. Because $\overline{g}_V^\ell$ is very small, not only $A_{LR}^0 = A_e$, $A_{FB}^{(0,\ell)}$, and $\mathcal{P}_\tau$, but also $A_{FB}^{(0,b)}$, $A_{FB}^{(0,c)}$, $A_{FB}^{(0,s)}$, and the hadronic asymmetries are mainly sensitive to $\overline{s}_\ell^2$. One finds that $\widehat{\kappa}_f$ ($f \neq b$) is almost independent of $(m_t, M_H)$, so that one can write

$$\overline{s}_\ell^2 \sim \widehat{s}_Z^2 + 0.00029 \ . \tag{10.34}$$

Thus, the asymmetries determine values of $\overline{s}_\ell^2$ and $\widehat{s}_Z^2$ almost independent of $m_t$, while the $\kappa$'s for the other schemes are $m_t$ dependent.

LEP 2 [41] has run at several energies above the $Z$-pole up to $\sim 209$ GeV. Measurements have been made of a number of observables, including the cross-sections for $e^+e^- \to f\bar{f}$ for $f = q, \mu^-, \tau^-$; the differential cross-sections and $A_{FB}$ for $\mu$ and $\tau$; $R$ and $A_{FB}$ for $b$ and $c$; $W$ branching ratios; and $WW$, $WW\gamma$, $ZZ$, single $W$, and single $Z$ cross-sections. They are in agreement with the SM predictions, with the exceptions of the total hadronic cross-section (1.7 $\sigma$ high), $R_b$ (2.1 $\sigma$ low), and $A_{FB}(b)$ (1.6 $\sigma$ low). Also, the SM Higgs has been excluded below 114.4 GeV [106].

The $Z$-boson properties are extracted assuming the SM expressions for the $\gamma$–$Z$ interference terms. These have also been tested experimentally by performing more general fits [107] to the LEP 1 and LEP 2 data. Assuming family universality this approach introduces three additional parameters relative to the standard fit [41], describing the $\gamma$–$Z$ interference contribution to the total hadronic and leptonic cross-sections, $j_{\text{had}}^{\text{tot}}$ and $j_\ell^{\text{tot}}$, and to the leptonic forward-backward asymmetry, $j_\ell^{\text{fb}}$. For example,

$$j_{\text{had}}^{\text{tot}} \sim g_V^\ell g_V^{\text{had}} = 0.277 \pm 0.065 \ , \tag{10.35}$$

which is in good agreement with the SM expectation [41] of $0.220^{+0.003}_{-0.014}$. Similarly, LEP data up to CM energies of 206 GeV were used to constrain the $\gamma$–$Z$ interference terms for the heavy quarks. The results for $j_b^{\text{tot}}$, $j_b^{\text{fb}}$, $j_c^{\text{tot}}$, and $j_c^{\text{fb}}$ were found in perfect agreement with the SM. These are valuable tests of the SM; but it should be cautioned that new physics is not expected to be described by this set of parameters, since (i) they do not account for extra interactions beyond the standard weak neutral-current, and (ii) the photonic amplitude remains fixed to its SM value.

Strong constraints on anomalous triple and quartic gauge couplings have been obtained at LEP 2 and at the Tevatron, as are described in the Particle Listings.

October 24, 2018 06:10

The left-right asymmetry in polarized Møller scattering $e^+e^- \to e^+e^-$ is being measured in the SLAC E158 experiment. A precision of better than $\pm 0.001$ in $\sin^2\theta_W$ at $Q^2 \sim 0.03$ GeV$^2$ is anticipated. The result of the first of three runs yields $\hat{s}_Z^2 = 0.2279 \pm 0.0032$ [108]. In a similar experiment and at about the same $Q^2$, Qweak at Jefferson Lab [109] will be able to measure $\sin^2\theta_W$ in polarized $ep$ scattering with a relative precision of 0.3%. These experiments will provide the most precise determinations of the weak mixing angle off the $Z$ peak and will be sensitive to various types of physics beyond the SM.

The Belle [110], CLEO [111], and BaBar [112] collaborations reported precise measurements of the flavor changing transition $b \to s\gamma$. The signal efficiencies (including the extrapolation to the full photon spectrum) depend on the bottom pole mass, $m_b$. We adjusted the Belle and BaBar results to agree with the $m_b$ value used by CLEO. In the case of CLEO, a 3.8% component from the model error of the signal efficiency is moved from the systematic error to the model error. The results for the branching fractions are then given by,

$$\mathcal{B}[\text{Belle}] = 3.05 \times 10^{-4} [1 \pm 0.158 \pm 0.124 \pm 0.202 \pm 0] , \quad (10.36a)$$

$$\mathcal{B}[\text{CLEO}] = 3.21 \times 10^{-4} [1 \pm 0.134 \pm 0.076 \pm 0.059 \pm 0.016] , \quad (10.36b)$$

$$\mathcal{B}[\text{BaBar}] = 3.86 \times 10^{-4} [1 \pm 0.090 \pm 0.093 \pm 0.074 \pm 0.016] , \quad (10.36c)$$

where the first two errors are the statistical and systematic uncertainties (taken uncorrelated). The third error (taken 100% correlated) accounts for the extrapolation from the finite photon energy cutoff (2.25 GeV, 2.0 GeV, and 2.1 GeV, respectively) to the full theoretical branching ratio [113]. The last error is from the correction for the $b \to d\gamma$ component which is common to CLEO and BaBar. It is advantageous [114] to normalize the result with respect to the semi-leptonic branching fraction, $\mathcal{B}(b \to Xe\nu) = 0.1064 \pm 0.0023$, yielding,

$$R = \frac{\mathcal{B}(b \to s\gamma)}{\mathcal{B}(b \to Xe\nu)} = (3.39 \pm 0.43 \pm 0.37) \times 10^{-3}. \quad (10.37)$$

In the fits we use the variable $\ln R = -5.69 \pm 0.17$ to assure an approximately Gaussian error [115]. We added an 11% theory uncertainty (excluding parametric errors such as from $\alpha_s$) in the SM prediction which is based on the next-to-leading order calculations of Refs. 114,116.

The present world average of the muon anomalous magnetic moment,

$$a_\mu^{\text{exp}} = \frac{g_\mu - 2}{2} = (1165920.37 \pm 0.78) \times 10^{-9} , \quad (10.38)$$

is dominated by the 1999 and 2000 data runs of the E821 collaboration at BNL [117]. The final 2001 data run is currently being analyzed. The QED contribution has been calculated to four loops (fully analytically to three loops), and the leading logarithms are included to five loops [118]. The estimated SM electroweak contribution [119–121], $a_\mu^{\rm EW} = (1.52 \pm 0.03) \times 10^{-9}$, which includes leading two-loop [120] and three-loop [121] corrections, is at the level of the current uncertainty. The limiting factor in the interpretation of the result is the uncertainty from the two-loop hadronic contribution [20], $a_\mu^{\rm had} = (69.63 \pm 0.72) \times 10^{-9}$, which has been obtained using $e^+ e^- \to$ hadrons cross-section data. The latter are dominated by the recently reanalyzed CMD 2 data [19]. This value suggests a 1.9 $\sigma$ discrepancy between Eq. (10.38) and the SM prediction. In an alternative analysis, the authors of Ref. 20 use $\tau$ decay data and isospin symmetry (CVC) to obtain instead $a_\mu^{\rm had} = (71.10 \pm 0.58) \times 10^{-9}$. This result implies no conflict (0.7 $\sigma$) with Eq. (10.38). Thus, there is also a discrepancy between the $2\pi$ spectral functions obtained from the two methods. For example, if one uses the $e^+ e^-$ data and CVC to predict the branching ratio for $\tau^- \to \nu_\tau \pi^- \pi^0$ decays one obtains $24.52 \pm 0.32\%$ [20] while the average of the measured branching ratios by DELPHI [122], ALEPH, CLEO, L3, and OPAL [20] yields $25.43 \pm 0.09\%$, which is 2.8 $\sigma$ higher. It is important to understand the origin of this difference and to obtain additional experimental information (*e.g.*, from the radiative return method [37]) . Fortunately, this problem is less pronounced as far as $a_\mu^{\rm had}$ is concerned: due to the suppression at large $s$ (from where the conflict originates) the difference is only 1.7 $\sigma$ (or 1.9 $\sigma$ if one adds the 4 $\pi$ channel which by itself is consistent between the two methods). Note also that a part of this difference is due to the older $e^+ e^-$ data [20], and the direct conflict between $\tau$ decay data and CMD 2 is less significant. Isospin violating corrections have been estimated in Ref. 123 and found to be under control. The largest effect is due to higher-order electroweak corrections [39] but introduces a negligible uncertainty [124]. In the following we view the 1.7 $\sigma$ difference as a fluctuation and average the results. An additional uncertainty is induced by the hadronic three-loop light-by-light scattering contribution [125], $a_\mu^{\rm LBLS} = (+0.83 \pm 0.19) \times 10^{-9}$, which was estimated within a form factor approach. The sign of this effect is opposite to the one quoted in the 2002 edition of this *Review*, and has subsequently been confirmed by two other groups [126]. Other hadronic effects at three-loop order contribute [127], $a_\mu^{\rm had}\left[\left(\frac{\alpha}{\pi}\right)^3\right] = (-1.00 \pm 0.06) \times 10^{-9}$. Correlations with the two-loop hadronic contribution and with $\Delta\alpha(M_Z)$ (see Sec. 10.2) were considered in Ref. 128, which also contains analytic results for the perturbative QCD contribution. The SM prediction is

$$a_\mu^{\rm theory} = (1165918.83 \pm 0.49) \times 10^{-9} , \qquad (10.39)$$

where the error is from the hadronic uncertainties excluding parametric ones such as from $\alpha_s$ and the heavy quark masses. We estimate its correlation with $\Delta\alpha(M_Z)$ as 21%. The small overall discrepancy between the experimental and theoretical values could be due to fluctuations or underestimates of the theoretical uncertainties. On the other hand, $g_\mu - 2$ is also affected by many types of new physics, such as supersymmetric models with large $\tan\beta$ and moderately light superparticle masses [129]. Thus, the deviation could also arise from physics beyond the SM.

*Note added:* After completion of this Section and the fits described here, the E821 collaboration announced its measurement on the anomalous magnetic moment of the negatively charged muon based on data taken in 2001 [130]. The result, $a_\mu^{\rm exp} = (1165921.4 \pm 0.8 \pm 0.3) \times 10^{-9}$, is consistent with the results on positive muons and appears to confirm the deviation. There also appeared two new evaluations [131,132] of $a_\mu^{\rm had}$. They are based on $e^+e^-$ data only and are generally in good agreement with each other and other $e^+e^-$ based analyzes. $\tau$ decay data are not used; it is argued [131] that CVC breaking effects (*e.g.*, through a relatively large mass difference between the $\rho^\pm$ and $\rho^0$ vector mesons) may be larger than expected. This may also be relevant in the context of the NuTeV discrepancy discussed above [131].

### 10.4. *W* and *Z* decays

The partial decay width for gauge bosons to decay into massless fermions $f_1\overline{f}_2$ is

$$\Gamma\left(W^+ \to e^+\nu_e\right) = \frac{G_F M_W^3}{6\sqrt{2}\pi} \approx 226.56 \pm 0.24 \text{ MeV}, \quad (10.47a)$$

$$\Gamma\left(W^+ \to u_i\overline{d}_j\right) = \frac{CG_F M_W^3}{6\sqrt{2}\pi}|V_{ij}|^2 \approx (707.1 \pm 0.7)\,|V_{ij}|^2 \text{ MeV}, \quad (10.47b)$$

$$\Gamma\left(Z \to \psi_i\overline{\psi}_i\right) = \frac{CG_F M_Z^3}{6\sqrt{2}\pi}\left[g_V^{i2} + g_A^{i2}\right] \quad (10.47c)$$

$$\approx \begin{cases} 300.4 \pm 0.2 \text{ MeV } (u\overline{u}), & 167.29 \pm 0.07 \text{ MeV } (\nu\overline{\nu}), \\ 383.2 \pm 0.2 \text{ MeV } (d\overline{d}), & 84.03 \pm 0.04 \text{ MeV } (e^+e^-), \\ 375.8 \mp 0.1 \text{ MeV } (b\overline{b}). \end{cases}$$

For leptons $C = 1$, while for quarks $C = 3\left(1 + \alpha_s(M_V)/\pi + 1.409\alpha_s^2/\pi^2 - 12.77\alpha_s^3/\pi^3\right)$, where the 3 is due to color and the factor in parentheses represents the universal part of the QCD corrections [133] for massless quarks [134]. The $Z \to f\overline{f}$ widths contain a number of additional corrections: universal (non-singlet) top quark mass contributions [135]; fermion mass effects and further QCD corrections proportional to $\widehat{m}_q^2(M_Z^2)$ [136] which are different for vector and axial-vector partial widths; and singlet contributions starting from two-loop order which are large, strongly top quark mass dependent, family universal, and flavor non-universal [137]. All QCD effects are known and included up to three-loop order. The QED factor $1 + 3\alpha q_f^2/4\pi$, as well as two-loop order $\alpha\alpha_s$ and $\alpha^2$ self-energy corrections [138] are also included. Working in the on-shell scheme, *i.e.*, expressing the widths in terms of $G_F M_{W,Z}^3$, incorporates the largest radiative corrections from the running QED coupling [44,139]. Electroweak corrections to the $Z$ widths are then incorporated by replacing $g_{V,A}^{i2}$ by $\overline{g}_{V,A}^{i2}$. Hence, in the on-shell scheme the $Z$ widths are proportional to $\rho_i \sim 1 + \rho_t$. The $\overline{\text{MS}}$ normalization accounts also for the leading electroweak corrections [48]. There is additional (negative) quadratic $m_t$ dependence in the $Z \to b\overline{b}$ vertex corrections [140] which causes $\Gamma(b\overline{b})$ to decrease with $m_t$. The dominant effect is to multiply $\Gamma(b\overline{b})$ by the vertex correction $1 + \delta\rho_{b\overline{b}}$, where $\delta\rho_{b\overline{b}} \sim 10^{-2}(-\frac{1}{2}\frac{m_t^2}{M_Z^2} + \frac{1}{5})$. In practice, the corrections are included in $\rho_b$ and $\kappa_b$, as discussed before.

For 3 fermion families the total widths are predicted to be

$$\Gamma_Z \approx 2.4968 \pm 0.0011 \text{ GeV} , \tag{10.48}$$
$$\Gamma_W \approx 2.0936 \pm 0.0022 \text{ GeV} . \tag{10.49}$$

We have assumed $\alpha_s(M_Z) = 0.1200$. An uncertainty in $\alpha_s$ of $\pm 0.0018$ introduces an additional uncertainty of 0.05% in the hadronic widths, corresponding to $\pm 0.9$ MeV in $\Gamma_Z$. These predictions are to be compared with the experimental results $\Gamma_Z = 2.4952 \pm 0.0023$ GeV [41] and $\Gamma_W = 2.124 \pm 0.041$ GeV (see the Particle Listings for more details).

**Table 10.4:** Principal $Z$-pole and other observables, compared with the SM predictions for the global best fit values $M_Z = 91.1874 \pm 0.0021$ GeV, $M_H = 113^{+56}_{-40}$ GeV, $m_t = 176.9 \pm 4.0$ GeV, $\alpha_s(M_Z) = 0.1213 \pm 0.0018$, and $\widehat{\alpha}(M_Z)^{-1} = 127.906 \pm 0.019$. The LEP averages of the ALEPH, DELPHI, L3, and OPAL results include common systematic errors and correlations [41]. The heavy flavor results of LEP and SLD are based on common inputs and correlated, as well [100]. $\overline{s}^2_\ell(A^{(0,q)}_{FB})$ is the effective angle extracted from the hadronic charge asymmetry, which has some correlation with $A^{(0,b)}_{FB}$ which is currently neglected. The values of $\Gamma(\ell^+\ell^-)$, $\Gamma(\text{had})$, and $\Gamma(\text{inv})$ are not independent of $\Gamma_Z$, the $R_\ell$, and $\sigma_{\text{had}}$. The $m_t$ values are from the lepton plus jets channel of the CDF [6] and DØ [9] run I data, respectively. Results from the other channels and all correlations are also included. The first $M_W$ value is from UA2, CDF, and DØ [141], while the second one is from LEP 2 [41]. The first $M_W$ and $M_Z$ are correlated, but the effect is negligible due to the tiny $M_Z$ error. The three values of $A_e$ are (i) from $A_{LR}$ for hadronic final states [101]; (ii) from $A_{LR}$ for leptonic final states and from polarized Bhabba scattering [103]; and (iii) from the angular distribution of the $\tau$ polarization. The two $A_\tau$ values are from SLD and the total $\tau$ polarization, respectively. $g^2_L$ and $g^2_R$ are from NuTeV [69] and have a very small ($-1.7\%$) residual anticorrelation. The older deep-inelastic scattering (DIS) results from CDHS [63], CHARM [64], and CCFR [65] are included, as well, but not shown in the Table. The world averages for $g^{\nu e}_{V,A}$ are dominated by the CHARM II [76] results, $g^{\nu e}_V = -0.035 \pm 0.017$ and $g^{\nu e}_A = -0.503 \pm 0.017$. The errors in $Q_W$, DIS, $b \to s\gamma$, and $g_\mu - 2$ are the total (experimental plus theoretical) uncertainties. The $\tau_\tau$ value is the $\tau$ lifetime world average computed by combining the direct measurements with values derived from the leptonic branching ratios [5]; the theory uncertainty is included in the SM prediction. In all other SM predictions, the uncertainty is from $M_Z$, $M_H$, $m_t$, $m_b$, $m_c$, $\widehat{\alpha}(M_Z)$, and $\alpha_s$, and their correlations have been accounted for. The SM errors in $\Gamma_Z$, $\Gamma(\text{had})$, $R_\ell$, and $\sigma_{\text{had}}$ are largely dominated by the uncertainty in $\alpha_s$.

| Quantity | Value | Standard Model | Pull |
| --- | --- | --- | --- |
| $m_t$ [GeV] | $176.1 \pm 7.4$ | $176.9 \pm 4.0$ | $-0.1$ |
|  | $180.1 \pm 5.4$ |  | $0.6$ |
| $M_W$ [GeV] | $80.454 \pm 0.059$ | $80.390 \pm 0.018$ | $1.1$ |
|  | $80.412 \pm 0.042$ |  | $0.5$ |
| $M_Z$ [GeV] | $91.1876 \pm 0.0021$ | $91.1874 \pm 0.0021$ | $0.1$ |
| $\Gamma_Z$ [GeV] | $2.4952 \pm 0.0023$ | $2.4972 \pm 0.0012$ | $-0.9$ |

**Table 10.4:** (continued)

| Quantity | Value | Standard Model | Pull |
|---|---|---|---|
| $\Gamma(\text{had})$ [GeV] | $1.7444 \pm 0.0020$ | $1.7435 \pm 0.0011$ | — |
| $\Gamma(\text{inv})$ [MeV] | $499.0 \pm 1.5$ | $501.81 \pm 0.13$ | — |
| $\Gamma(\ell^+\ell^-)$ [MeV] | $83.984 \pm 0.086$ | $84.024 \pm 0.025$ | — |
| $\sigma_{\text{had}}$ [nb] | $41.541 \pm 0.037$ | $41.472 \pm 0.009$ | $1.9$ |
| $R_e$ | $20.804 \pm 0.050$ | $20.750 \pm 0.012$ | $1.1$ |
| $R_\mu$ | $20.785 \pm 0.033$ | $20.751 \pm 0.012$ | $1.0$ |
| $R_\tau$ | $20.764 \pm 0.045$ | $20.790 \pm 0.018$ | $-0.7$ |
| $R_b$ | $0.21638 \pm 0.00066$ | $0.21564 \pm 0.00014$ | $1.1$ |
| $R_c$ | $0.1720 \pm 0.0030$ | $0.17233 \pm 0.00005$ | $-0.1$ |
| $A_{FB}^{(0,e)}$ | $0.0145 \pm 0.0025$ | $0.01626 \pm 0.00025$ | $-0.7$ |
| $A_{FB}^{(0,\mu)}$ | $0.0169 \pm 0.0013$ | | $0.5$ |
| $A_{FB}^{(0,\tau)}$ | $0.0188 \pm 0.0017$ | | $1.5$ |
| $A_{FB}^{(0,b)}$ | $0.0997 \pm 0.0016$ | $0.1032 \pm 0.0008$ | $-2.2$ |
| $A_{FB}^{(0,c)}$ | $0.0706 \pm 0.0035$ | $0.0738 \pm 0.0006$ | $-0.9$ |
| $A_{FB}^{(0,s)}$ | $0.0976 \pm 0.0114$ | $0.1033 \pm 0.0008$ | $-0.5$ |
| $\bar{s}_\ell^2(A_{FB}^{(0,q)})$ | $0.2324 \pm 0.0012$ | $0.23149 \pm 0.00015$ | $0.8$ |
| $A_e$ | $0.15138 \pm 0.00216$ | $0.1472 \pm 0.0011$ | $1.9$ |
| | $0.1544 \pm 0.0060$ | | $1.2$ |
| | $0.1498 \pm 0.0049$ | | $0.5$ |
| $A_\mu$ | $0.142 \pm 0.015$ | | $-0.4$ |
| $A_\tau$ | $0.136 \pm 0.015$ | | $-0.8$ |
| | $0.1439 \pm 0.0043$ | | $-0.8$ |
| $A_b$ | $0.925 \pm 0.020$ | $0.9347 \pm 0.0001$ | $-0.5$ |
| $A_c$ | $0.670 \pm 0.026$ | $0.6678 \pm 0.0005$ | $0.1$ |
| $A_s$ | $0.895 \pm 0.091$ | $0.9357 \pm 0.0001$ | $-0.4$ |
| $g_L^2$ | $0.30005 \pm 0.00137$ | $0.30397 \pm 0.00023$ | $-2.9$ |
| $g_R^2$ | $0.03076 \pm 0.00110$ | $0.03007 \pm 0.00003$ | $0.6$ |
| $g_V^{\nu e}$ | $-0.040 \pm 0.015$ | $-0.0397 \pm 0.0003$ | $-0.1$ |
| $g_A^{\nu e}$ | $-0.507 \pm 0.014$ | $-0.5065 \pm 0.0001$ | $0.0$ |
| $Q_W(\text{Cs})$ | $-72.69 \pm 0.48$ | $-73.19 \pm 0.03$ | $1.0$ |
| $Q_W(\text{Tl})$ | $-116.6 \pm 3.7$ | $-116.81 \pm 0.04$ | $0.1$ |
| $\frac{\Gamma(b \to s\gamma)}{\Gamma(b \to Xe\nu)}$ | $3.39^{+0.62}_{-0.54} \times 10^{-3}$ | $(3.23 \pm 0.09) \times 10^{-3}$ | $0.3$ |
| $\frac{1}{2}(g_\mu - 2 - \frac{\alpha}{\pi})$ | $4510.64 \pm 0.92$ | $4509.13 \pm 0.10$ | $1.6$ |
| $\tau_\tau$ [fs] | $290.92 \pm 0.55$ | $291.83 \pm 1.81$ | $-0.4$ |

## 10.5. Experimental results

The values of the principal $Z$-pole observables are listed in Table 10.4, along with the SM predictions for $M_Z = 91.1874 \pm 0.0021$ GeV, $M_H = 113^{+56}_{-40}$ GeV, $m_t = 176.9 \pm 4.0$ GeV, $\alpha_s(M_Z) = 0.1213 \pm 0.0018$, and $\widehat{\alpha}(M_Z)^{-1} = 127.906 \pm 0.019$ ($\Delta\alpha^{(5)}_{\text{had}} \approx 0.02801 \pm 0.00015$). The values and predictions of $M_W$ [41,141]; $m_t$ [6,9]; the $Q_W$ for cesium [81] and thallium [82]; deep inelastic [69] and $\nu_\mu$-$e$ scattering [74–76]; the $b \to s\gamma$ observable [110–112]; the muon anomalous magnetic moment [117]; and the $\tau$ lifetime are also listed. The values of $M_W$ and $m_t$ differ from those in the Particle Listings because they include recent preliminary results. The agreement is excellent. Only $g^2_L$ from NuTeV is currently showing a large (2.9 $\sigma$) deviation. In addition, the hadronic peak cross-section, $\sigma_{\text{had}}$, and the $A^0_{LR}$ from hadronic final states differ by 1.9 $\sigma$. On the other hand, $A^{(0,b)}_{FB}$ (2.2 $\sigma$) and $g_\mu - 2$ (1.6 $\sigma$, see Sec. 10.3) both moved closer to the SM predictions by about one standard deviation compared to the 2002 edition of this *Review*, while $M_W$ (LEP 2) has moved closer by 0.8 $\sigma$. Observables like $R_b = \Gamma(b\bar{b})/\Gamma(\text{had})$, $R_c = \Gamma(c\bar{c})/\Gamma(\text{had})$, and the combined value for $M_W$ which showed significant deviations in the past, are now in reasonable agreement. In particular, $R_b$ whose measured value deviated as much as 3.7 $\sigma$ from the SM prediction is now only 1.1 $\sigma$ (0.34%) high.

$A_b$ can be extracted from $A^{(0,b)}_{FB}$ when $A_e = 0.1501 \pm 0.0016$ is taken from a fit to leptonic asymmetries (using lepton universality). The result, $A_b = 0.886 \pm 0.017$, is 2.9 $\sigma$ below the SM prediction†, and also 1.5 $\sigma$ below $A_b = 0.925 \pm 0.020$ obtained from $A^{FB}_{LR}(b)$ at SLD. Thus, it appears that at least some of the problem in $A^{(0,b)}_{FB}$ is experimental. Note, however, that the uncertainty in $A^{(0,b)}_{FB}$ is strongly statistics dominated. The combined value, $A_b = 0.902 \pm 0.013$ deviates by 2.5 $\sigma$. It would be extremely difficult to account for this 3.5% deviation by new physics radiative corrections since an order of 20% correction to $\widehat{\kappa}_b$ would be necessary to account for the central value of $A_b$. If this deviation is due to new physics, it is most likely of tree-level type affecting preferentially the third generation. Examples include the decay of a scalar neutrino resonance [142], mixing of the $b$ quark with heavy exotics [143], and a heavy $Z'$ with family-nonuniversal couplings [144]. It is difficult, however, to simultaneously account for $R_b$, which has been measured on the $Z$ peak and off-peak [145] at LEP 1. An average of $R_b$ measurements at LEP 2 at energies between 133 and 207 GeV is 2.1 $\sigma$ below the SM prediction, while $A^{(b)}_{FB}$(LEP 2) is 1.6 $\sigma$ low.

---

† Alternatively, one can use $A_\ell = 0.1481 \pm 0.0027$, which is from LEP alone and in excellent agreement with the SM, and obtain $A_b = 0.898 \pm 0.022$ which is 1.7 $\sigma$ low. This illustrates that some of the discrepancy is related to the one in $A_{LR}$.

The left-right asymmetry, $A_{LR}^0 = 0.15138 \pm 0.00216$ [101], based on all hadronic data from 1992–1998 differs 1.9 $\sigma$ from the SM expectation of $0.1472 \pm 0.0011$. The combined value of $A_\ell = 0.1513 \pm 0.0021$ from SLD (using lepton-family universality and including correlations) is also 1.9 $\sigma$ above the SM prediction; but there is now experimental agreement between this SLD value and the LEP value, $A_\ell = 0.1481 \pm 0.0027$, obtained from a fit to $A_{FB}^{(0,\ell)}$, $A_e(\mathcal{P}_\tau)$, and $A_\tau(\mathcal{P}_\tau)$, again assuming universality.

Despite these discrepancies the goodness of the fit to all data is excellent with a $\chi^2$/d.o.f. = 45.5/45. The probability of a larger $\chi^2$ is 45%. The observables in Table 10.4, as well as some other less precise observables, are used in the global fits described below. The correlations on the LEP lineshape and $\tau$ polarization, the LEP/SLD heavy flavor observables, the SLD lepton asymmetries, the deep inelastic and $\nu$-$e$ scattering observables, and the $m_t$ measurements, are included. The theoretical correlations between $\Delta\alpha_{had}^{(5)}$ and $g_\mu - 2$, and between the charm and bottom quark masses, are also accounted for.

**Table 10.5:** Values of $\hat{s}_Z^2$, $s_W^2$, $\alpha_s$, and $M_H$ [in GeV] for various (combinations of) observables. Unless indicated otherwise, the top quark mass, $m_t = 177.9 \pm 4.4$ GeV, is used as an additional constraint in the fits. The (†) symbol indicates a fixed parameter.

| Data | $\hat{s}_Z^2$ | $s_W^2$ | $\alpha_s(M_Z)$ | $M_H$ |
|---|---|---|---|---|
| All data | 0.23120(15) | 0.2228(4) | 0.1213(18) | $113^{+56}_{-40}$ |
| All indirect (no $m_t$) | 0.23116(17) | 0.2229(4) | 0.1213(18) | $79^{+95}_{-38}$ |
| Z pole (no $m_t$) | 0.23118(17) | 0.2231(6) | 0.1197(28) | $79^{+94}_{-38}$ |
| LEP 1 (no $m_t$) | 0.23148(20) | 0.2237(7) | 0.1210(29) | $140^{+192}_{-74}$ |
| SLD + $M_Z$ | 0.23067(28) | 0.2217(6) | 0.1213 (†) | $43^{+38}_{-23}$ |
| $A_{FB}^{(b,c)}$ + $M_Z$ | 0.23185(28) | 0.2244(8) | 0.1213 (†) | $408^{+317}_{-179}$ |
| $M_W$ + $M_Z$ | 0.23089(37) | 0.2221(8) | 0.1213 (†) | $67^{+77}_{-45}$ |
| $M_Z$ | 0.23117(15) | 0.2227(5) | 0.1213 (†) | 117 (†) |
| DIS (isoscalar) | 0.2359(16) | 0.2274(16) | 0.1213 (†) | 117 (†) |
| $Q_W$ (APV) | 0.2292(19) | 0.2207(19) | 0.1213 (†) | 117 (†) |
| polarized Møller | 0.2292(42) | 0.2207(43) | 0.1213 (†) | 117 (†) |
| elastic $\nu_\mu(\overline{\nu_\mu})e$ | 0.2305(77) | 0.2220(77) | 0.1213 (†) | 117 (†) |
| SLAC $eD$ | 0.222(18) | 0.213(19) | 0.1213 (†) | 117 (†) |
| elastic $\nu_\mu(\overline{\nu_\mu})p$ | 0.211(33) | 0.203(33) | 0.1213 (†) | 117 (†) |

The data allow a simultaneous determination of $M_H$, $m_t$, $\sin^2\theta_W$, and the strong coupling $\alpha_s(M_Z)$. ($\widehat{m}_c$, $\widehat{m}_b$, and $\Delta\alpha_{\text{had}}^{(5)}$ are also allowed to float in the fits, subject to the theoretical constraints [5,18] described in Sec. 10.1–Sec. 10.2. These are correlated with $\alpha_s$.) $\alpha_s$ is determined mainly from $R_\ell$, $\Gamma_Z$, $\sigma_{\text{had}}$, and $\tau_\tau$ and is only weakly correlated with the other variables (except for a 10% correlation with $\widehat{m}_c$). The global fit to all data, including the CDF/DØ average, $m_t = 177.9 \pm 4.4$ GeV, yields

$$\begin{aligned}
M_H &= 113^{+56}_{-40} \text{ GeV}, \\
m_t &= 176.9 \pm 4.0 \text{ GeV}, \\
\widehat{s}_Z^2 &= 0.23120 \pm 0.00015, \\
\alpha_s(M_Z) &= 0.1213 \pm 0.0018.
\end{aligned} \qquad (10.50)$$

In the on-shell scheme one has $s_W^2 = 0.22280 \pm 0.00035$, the larger error due to the stronger sensitivity to $m_t$, while the corresponding effective angle is related by Eq. (10.34), i.e., $\overline{s}_\ell^2 = 0.23149 \pm 0.00015$. The $m_t$ pole mass corresponds to $\widehat{m}_t(\widehat{m}_t) = 166.8 \pm 3.8$ GeV. In all fits, the errors include full statistical, systematic, and theoretical uncertainties. The $\widehat{s}_Z^2$ ($\overline{s}_\ell^2$) error reflects the error on $\overline{s}_f^2 = 0.23150 \pm 0.00016$ from a fit to the $Z$-pole asymmetries.

The weak mixing angle can be determined from $Z$-pole observables, $M_W$, and from a variety of neutral-current processes spanning a very wide $Q^2$ range. The results (for the older low-energy neutral-current data see [42,43]) shown in Table 10.5 are in reasonable agreement with each other, indicating the quantitative success of the SM. The largest discrepancy is the value $\widehat{s}_Z^2 = 0.2358 \pm 0.0016$ from DIS which is 2.9 $\sigma$ above the value $0.23120 \pm 0.00015$ from the global fit to all data. Similarly, $\widehat{s}_Z^2 = 0.23185 \pm 0.00028$ from the forward-backward asymmetries into bottom and charm quarks, and $\widehat{s}_Z^2 = 0.23067 \pm 0.00028$ from the SLD asymmetries (both when combined with $M_Z$) are 2.3 $\sigma$ high and 1.9 $\sigma$ low, respectively.

The extracted $Z$-pole value of $\alpha_s(M_Z)$ is based on a formula with negligible theoretical uncertainty ($\pm 0.0005$ in $\alpha_s(M_Z)$) if one assumes the exact validity of the SM. One should keep in mind, however, that this value, $\alpha_s = 0.1197 \pm 0.0028$, is very sensitive to such types of new physics as non-universal vertex corrections. In contrast, the value derived from $\tau$ decays, $\alpha_s(M_Z) = 0.1221^{+0.0026}_{-0.0023}$ [5], is theory dominated but less sensitive to new physics. The former is mainly due to the larger value of $\alpha_s(m_\tau)$, but just as the hadronic $Z$-width the $\tau$ lifetime is fully inclusive and can be computed reliably within the operator product expansion. The two values are in excellent agreement with each other. They are also in perfect agreement with other recent values, such as $0.1202 \pm 0.0049$ from jet-event shapes at LEP [146], and $0.121 \pm 0.003$ [147] from the most recent lattice calculation of the $\Upsilon$ spectrum. For more details and other determinations, see our Section 9 on "Quantum Chromodynamics" in this *Review*.

The data indicate a preference for a small Higgs mass. There is a strong correlation between the quadratic $m_t$ and logarithmic $M_H$

terms in $\hat{\rho}$ in all of the indirect data except for the $Z \to b\bar{b}$ vertex. Therefore, observables (other than $R_b$) which favor $m_t$ values higher than the Tevatron range favor lower values of $M_H$. This effect is enhanced by $R_b$, which has little direct $M_H$ dependence but favors the lower end of the Tevatron $m_t$ range. $M_W$ has additional $M_H$ dependence through $\Delta \hat{r}_W$ which is not coupled to $m_t^2$ effects. The strongest individual pulls toward smaller $M_H$ are from $M_W$ and $A_{LR}^0$, while $A_{FB}^{(0b)}$ and the NuTeV results favor high values. The difference in $\chi^2$ for the global fit is $\Delta \chi^2 = \chi^2(M_H = 1000 \text{ GeV}) - \chi^2_{\min} = 34.6$. Hence, the data favor a small value of $M_H$, as in supersymmetric extensions of the SM. The central value of the global fit result, $M_H = 113^{+56}_{-40}$ GeV, is slightly below the direct lower bound, $M_H \geq 114.4$ GeV (95% CL) [106].

The 90% central confidence range from all precision data is

$$53 \text{ GeV} \leq M_H \leq 213 \text{ GeV} .$$

Including the results of the direct searches as an extra contribution to the likelihood function drives the 95% upper limit to $M_H \leq 241$ GeV. As two further refinements, we account for (i) theoretical uncertainties from uncalculated higher order contributions by allowing the $T$ parameter (see next subsection) subject to the constraint $T = 0 \pm 0.02$, (ii) the $M_H$ dependence of the correlation matrix which gives slightly more weight to lower Higgs masses [148]. The resulting limits at 95 (90, 99)% CL are

$$M_H \leq 246 \ (217, \ 311) \text{ GeV} ,$$

respectively. The extraction of $M_H$ from the precision data depends strongly on the value used for $\alpha(M_Z)$. Upper limits, however, are more robust due to two compensating effects: the older results indicated more QED running and were less precise, yielding $M_H$ distributions which were broader with centers shifted to smaller values. The hadronic contribution to $\alpha(M_Z)$ is correlated with $g_\mu - 2$ (see Sec. 10.3). The measurement of the latter is higher than the SM prediction, and its inclusion in the fit favors a larger $\alpha(M_Z)$ and a lower $M_H$ (by 4 GeV).

One can also carry out a fit to the indirect data alone, *i.e.*, without including the constraint, $m_t = 177.9 \pm 4.4$ GeV, obtained by CDF and DØ. (The indirect prediction is for the $\overline{\text{MS}}$ mass, $\hat{m}_t(\hat{m}_t) = 162.5^{+9.2}_{-6.9}$ GeV, which is in the end converted to the pole mass). One obtains $m_t = 172.4^{+9.8}_{-7.3}$ GeV, with little change in the $\sin^2 \theta_W$ and $\alpha_s$ values, in remarkable agreement with the direct CDF/DØ average. The relations between $M_H$ and $m_t$ for various observables are shown in Fig. 10.1.

Using $\alpha(M_Z)$ and $\hat{s}_Z^2$ as inputs, one can predict $\alpha_s(M_Z)$ assuming grand unification. One predicts [149] $\alpha_s(M_Z) = 0.130 \pm 0.001 \pm 0.01$ for the simplest theories based on the minimal supersymmetric extension of the SM, where the first (second) uncertainty is from the inputs (thresholds). This is slightly larger, but consistent with the experimental $\alpha_s(M_Z) = 0.1213 \pm 0.0018$ from the Z lineshape and the $\tau$ lifetime, as well as with other determinations. Non-supersymmetric unified theories predict the low value $\alpha_s(M_Z) = 0.073 \pm 0.001 \pm 0.001$. See also the note on "Low-Energy Supersymmetry" in the Particle Listings.

**Figure 10.1:** One-standard-deviation (39.35%) uncertainties in $M_H$ as a function of $m_t$ for various inputs, and the 90% CL region ($\Delta\chi^2 = 4.605$) allowed by all data. $\alpha_s(M_Z) = 0.120$ is assumed except for the fits including the Z-lineshape data. The 95% direct lower limit from LEP 2 is also shown.

One can also determine the radiative correction parameters $\Delta r$: from the global fit one obtains $\Delta r = 0.0347 \pm 0.0011$ and $\Delta \widehat{r}_W = 0.06981 \pm 0.00032$. $M_W$ measurements [41,141] (when combined with $M_Z$) are equivalent to measurements of $\Delta r = 0.0326 \pm 0.0021$, which is 1.2 $\sigma$ below the result from all indirect data, $\Delta r = 0.0355 \pm 0.0013$. Fig. 10.2 shows the 1 $\sigma$ contours in the $M_W - m_t$ plane from the direct and indirect determinations, as well as the combined 90% CL region. The indirect determination uses $M_Z$ from LEP 1 as input, which is defined assuming an s-dependent decay width. $M_W$ then corresponds to the s-dependent width definition, as well, and can be directly compared with the results from the Tevatron and LEP 2 which have been obtained using the same definition. The difference to a constant width definition is formally only of $\mathcal{O}(\alpha^2)$, but is strongly enhanced since the decay channels add up coherently. It is about 34 MeV for $M_Z$ and 27 MeV for $M_W$. The residual difference between working consistently with one or the other definition is about 3 MeV, *i.e.*, of typical size for non-enhanced $\mathcal{O}(\alpha^2)$ corrections [54,55].

**Figure 10.2:** One-standard-deviation (39.35%) region in $M_W$ as a function of $m_t$ for the direct and indirect data, and the 90% CL region ($\Delta\chi^2 = 4.605$) allowed by all data. The SM prediction as a function of $M_H$ is also indicated. The widths of the $M_H$ bands reflect the theoretical uncertainty from $\alpha(M_Z)$.

Most of the parameters relevant to $\nu$-hadron, $\nu$-e, e-hadron, and $e^+e^-$ processes are determined uniquely and precisely from the data in "model-independent" fits (*i.e.*, fits which allow for an arbitrary electroweak gauge theory). The values for the parameters defined in Eqs. (10.11)–(10.13) are given in Table 10.6 along with the predictions of the SM. The agreement is reasonable, except for the values of $g_L^2$ and $\epsilon_L(u,d)$, which reflect the discrepancy in the recent NuTeV results.

**Table 10.6:** Values of the model-independent neutral-current parameters, compared with the SM predictions for the global best fit values $M_Z = 91.1874 \pm 0.0021$ GeV, $M_H = 113^{+56}_{-40}$ GeV, $m_t = 176.9 \pm 4.0$ GeV, $\alpha_s(M_Z) = 0.1213 \pm 0.0018$, and $\widehat{\alpha}(M_Z)^{-1} = 127.906 \pm 0.019$. There is a second $g^{\nu e}_{V,A}$ solution, given approximately by $g^{\nu e}_V \leftrightarrow g^{\nu e}_A$, which is eliminated by $e^+e^-$ data under the assumption that the neutral current is dominated by the exchange of a single $Z$. The $\epsilon_L$, as well as the $\epsilon_R$, are strongly correlated and non-Gaussian, so that for implementations we recommend the parametrization using $g_i$ and $\theta_i = \tan^{-1}[\epsilon_i(u)/\epsilon_i(d)]$, $i = L$ or $R$. $\theta_R$ is only weakly correlated with the $g_i$, while the correlation coefficient between $\theta_R$ and $\theta_L$ is 0.27.

| Quantity | Experimental Value | SM | Correlation | | |
|---|---|---|---|---|---|
| $\epsilon_L(u)$ | $0.326 \pm 0.013$ | 0.3460(2) | | | |
| $\epsilon_L(d)$ | $-0.441 \pm 0.010$ | $-0.4292(1)$ | non- | | |
| $\epsilon_R(u)$ | $-0.175^{+0.013}_{-0.004}$ | $-0.1551(1)$ | Gaussian | | |
| $\epsilon_R(d)$ | $-0.022^{+0.072}_{-0.047}$ | 0.0776 | | | |
| $g_L^2$ | $0.3005 \pm 0.0012$ | 0.3040(2) | $-0.11$ | $-0.21$ | $-0.01$ |
| $g_R^2$ | $0.0311 \pm 0.0010$ | 0.0301 | | $-0.02$ | $-0.03$ |
| $\theta_L$ | $2.51 \pm 0.033$ | 2.4631(1) | | | 0.26 |
| $\theta_R$ | $4.59^{+0.41}_{-0.28}$ | 5.1765 | | | |
| $g_V^{\nu e}$ | $-0.040 \pm 0.015$ | $-0.0397(3)$ | | | $-0.05$ |
| $g_A^{\nu e}$ | $-0.507 \pm 0.014$ | $-0.5065(1)$ | | | |
| $C_{1u} + C_{1d}$ | $0.148 \pm 0.004$ | 0.1529(1) | 0.95 | $-0.55$ | $-0.26$ |
| $C_{1u} - C_{1d}$ | $-0.597 \pm 0.061$ | $-0.5299(4)$ | | $-0.57$ | $-0.27$ |
| $C_{2u} + C_{2d}$ | $0.62 \pm 0.80$ | $-0.0095$ | | | $-0.38$ |
| $C_{2u} - C_{2d}$ | $-0.07 \pm 0.12$ | $-0.0623(6)$ | | | |

(The $\nu$-hadron results without the new NuTeV data can be found in the previous editions of this *Review*.). The off $Z$-pole $e^+e^-$ results are difficult to present in a model-independent way because $Z$-propagator effects are non-negligible at TRISTAN, PETRA, PEP, and LEP 2 energies. However, assuming $e$-$\mu$-$\tau$ universality, the low-energy lepton asymmetries imply [98] $4(g_A^e)^2 = 0.99 \pm 0.05$, in good agreement with the SM prediction $\simeq 1$.

The results presented here are generally in reasonable agreement with the ones obtained by the LEP Electroweak Working Group [41]. We obtain higher best fit values for $\alpha_s$ and a higher and slightly more precise $M_H$. We trace most of the differences to be due to (i) the inclusion of recent higher order radiative corrections, in particular, the leading $\mathcal{O}(\alpha_s^4)$ contribution to hadronic $Z$ decays [150]; (ii) a different evaluation of $\alpha(M_Z)$ [18]; (iii) slightly different data sets (such as the recent DØ $m_t$ value); and (iv) scheme dependences. Taking into account these differences, the agreement is excellent.

## 10.6. Constraints on new physics

The $Z$-pole, $W$ mass, and neutral-current data can be used to search for and set limits on deviations from the SM. In particular, the combination of these indirect data with the direct CDF and DØ average for $m_t$ allows one to set stringent limits on new physics. We will mainly discuss the effects of exotic particles (with heavy masses $M_{\text{new}} \gg M_Z$ in an expansion in $M_Z/M_{\text{new}}$) on the gauge boson self-energies. (Brief remarks are made on new physics which is not of this type.) Most of the effects on precision measurements can be described by three gauge self-energy parameters $S$, $T$, and $U$. We will define these, as well as related parameters, such as $\rho_0$, $\epsilon_i$, and $\widehat{\epsilon}_i$, to arise from new physics only. *I.e.*, they are equal to zero ($\rho_0 = 1$) exactly in the SM, and do not include any contributions from $m_t$ or $M_H$, which are treated separately. Our treatment differs from most of the original papers.

Many extensions of the SM are described by the $\rho_0$ parameter,

$$\rho_0 \equiv M_W^2 / \left( M_Z^2 \, \widehat{c}_Z^2 \, \widehat{\rho} \right) , \qquad (10.51)$$

which describes new sources of SU(2) breaking that cannot be accounted for by the SM Higgs doublet or $m_t$ effects. In the presence of $\rho_0 \neq 1$, Eq. (10.51) generalizes Eq. (10.8$b$) while Eq. (10.8$a$) remains unchanged. Provided that the new physics which yields $\rho_0 \neq 1$ is a small perturbation which does not significantly affect the radiative corrections, $\rho_0$ can be regarded as a phenomenological parameter which multiplies $G_F$ in Eqs. (10.11)–(10.13), (10.28), and $\Gamma_Z$ in Eq. (10.47). There is enough data to determine $\rho_0$, $M_H$, $m_t$, and $\alpha_s$, simultaneously. From the global fit,

$$\rho_0 = 0.9998^{+0.0008}_{-0.0005} , \qquad (10.52)$$
$$114.4 \text{ GeV} < M_H < 193 \text{ GeV} , \qquad (10.53)$$
$$m_t = 178.0 \pm 4.1 \text{ GeV} , \qquad (10.54)$$
$$\alpha_s (M_Z) = 0.1214 \pm 0.0018 , \qquad (10.55)$$

where the lower limit on $M_H$ is the direct search bound. (If the direct limit is ignored one obtains $M_H = 66^{+86}_{-30}$ GeV and $\rho_0 = 0.9993^{+0.0010}_{-0.0008}$.) The error bar in Eq. (10.52) is highly asymmetric: at the 2 $\sigma$ level one has $\rho_0 = 0.9998^{+0.0025}_{-0.0010}$ and $M_H < 664$ GeV. Clearly, in the presence of $\rho_0$ upper limits on $M_H$ become much weaker. The result in Eq. (10.52) is in remarkable agreement with the SM expectation, $\rho_0 = 1$. It can be used to constrain higher-dimensional Higgs representations to have vacuum expectation values of less than a few percent of those of the doublets. Indeed, the relation between $M_W$ and $M_Z$ is modified if there are Higgs multiplets with weak isospin $> 1/2$ with significant vacuum expectation values. In order to calculate to higher orders in such theories one must define a set of four fundamental renormalized parameters which one may conveniently choose to be $\alpha$, $G_F$, $M_Z$, and $M_W$, since $M_W$ and $M_Z$ are directly measurable. Then $\hat{s}_Z^2$ and $\rho_0$ can be considered dependent parameters.

Eq. (10.52) can also be used to constrain other types of new physics. For example, non-degenerate multiplets of heavy fermions or scalars break the vector part of weak SU(2) and lead to a decrease in the value of $M_Z/M_W$. A non-degenerate SU(2) doublet $\binom{f_1}{f_2}$ yields a positive contribution to $\rho_0$ [151] of

$$\frac{CG_F}{8\sqrt{2}\pi^2}\Delta m^2 \;, \tag{10.56}$$

where

$$\Delta m^2 \equiv m_1^2 + m_2^2 - \frac{4m_1^2 m_2^2}{m_1^2 - m_2^2}\ln\frac{m_1}{m_2} \geq (m_1 - m_2)^2 \;, \tag{10.57}$$

and $C = 1$ (3) for color singlets (triplets). Thus, in the presence of such multiplets, one has

$$\frac{3G_F}{8\sqrt{2}\pi^2}\sum_i \frac{C_i}{3}\Delta m_i^2 = \rho_0 - 1 \;, \tag{10.58}$$

where the sum includes fourth-family quark or lepton doublets, $\binom{t'}{b'}$ or $\binom{E^0}{E^-}$, and scalar doublets such as $\binom{\tilde{t}}{\tilde{b}}$ in Supersymmetry (in the absence of $L - R$ mixing). This implies

$$\sum_i \frac{C_i}{3}\Delta m_i^2 \leq (85~\text{GeV})^2 \tag{10.59}$$

at 95% CL. The corresponding constraints on non-degenerate squark and slepton doublets are even stronger, $\sum_i C_i\Delta m_i^2/3 \leq (59~\text{GeV})^2$. This is due to the MSSM Higgs mass bound, $m_{h^0} < 150$ GeV, and the very strong correlation between $m_{h^0}$ and $\rho_0$ (79%).

Non-degenerate multiplets usually imply $\rho_0 > 1$. Similarly, heavy $Z'$ bosons decrease the prediction for $M_Z$ due to mixing and generally lead to $\rho_0 > 1$ [152]. On the other hand, additional Higgs doublets which participate in spontaneous symmetry breaking [153], heavy lepton doublets involving Majorana neutrinos [154], and the vacuum expectation values of Higgs triplets or higher-dimensional representations can contribute to $\rho_0$ with either sign. Allowing for the presence of heavy degenerate chiral multiplets (the $S$ parameter, to be discussed below) affects the determination of $\rho_0$ from the data, at present leading to a smaller value (for fixed $M_H$).

A number of authors [155–160] have considered the general effects on neutral-current and $Z$ and $W$ boson observables of various types of heavy (*i.e.*, $M_{\text{new}} \gg M_Z$) physics which contribute to the $W$ and $Z$ self-energies but which do not have any direct coupling to the ordinary fermions. In addition to non-degenerate multiplets, which break the vector part of weak SU(2), these include heavy degenerate multiplets of chiral fermions which break the axial generators. The effects of one degenerate chiral doublet are small, but in Technicolor theories there may be many chiral doublets and therefore significant effects [155].

Such effects can be described by just three parameters, $S$, $T$, and $U$ at the (electroweak) one-loop level. (Three additional parameters are needed if the new physics scale is comparable to $M_Z$ [161].) $T$ is proportional to the difference between the $W$ and $Z$ self-energies at $Q^2 = 0$ (*i.e.*, vector SU(2)-breaking), while $S$ ($S + U$) is associated with the difference between the $Z$ ($W$) self-energy at $Q^2 = M_{Z,W}^2$ and $Q^2 = 0$ (axial SU(2)-breaking). Denoting the contributions of new physics to the various self-energies by $\Pi_{ij}^{\text{new}}$, we have

$$\widehat{\alpha}(M_Z) T \equiv \frac{\Pi_{WW}^{\text{new}}(0)}{M_W^2} - \frac{\Pi_{ZZ}^{\text{new}}(0)}{M_Z^2}, \tag{10.60a}$$

$$\frac{\widehat{\alpha}(M_Z)}{4\widehat{s}_Z^2 \widehat{c}_Z^2} S \equiv \frac{\Pi_{ZZ}^{\text{new}}(M_Z^2) - \Pi_{ZZ}^{\text{new}}(0)}{M_Z^2}$$
$$- \frac{\widehat{c}_Z^2 - \widehat{s}_Z^2}{\widehat{c}_Z \widehat{s}_Z} \frac{\Pi_{Z\gamma}^{\text{new}}(M_Z^2)}{M_Z^2} - \frac{\Pi_{\gamma\gamma}^{\text{new}}(M_Z^2)}{M_Z^2}, \tag{10.60b}$$

$$\frac{\widehat{\alpha}(M_Z)}{4\widehat{s}_Z^2}(S+U) \equiv \frac{\Pi_{WW}^{\text{new}}(M_W^2) - \Pi_{WW}^{\text{new}}(0)}{M_W^2}$$
$$- \frac{\widehat{c}_Z}{\widehat{s}_Z} \frac{\Pi_{Z\gamma}^{\text{new}}(M_Z^2)}{M_Z^2} - \frac{\Pi_{\gamma\gamma}^{\text{new}}(M_Z^2)}{M_Z^2}. \tag{10.60c}$$

$S$, $T$, and $U$ are defined with a factor proportional to $\widehat{\alpha}$ removed, so that they are expected to be of order unity in the presence of new physics. In the $\overline{\text{MS}}$ scheme as defined in Ref. 46, the last two terms in Eq. (10.60*b*) and Eq. (10.60*c*) can be omitted (as was done in some earlier editions of this *Review*). They are related to other parameters ($S_i$, $h_i$, $\widehat{\epsilon}_i$) defined in [46,156,157] by

$$T = h_V = \widehat{\epsilon}_1 / \alpha,$$
$$S = h_{AZ} = S_Z = 4\widehat{s}_Z^2 \widehat{\epsilon}_3 / \alpha,$$
$$U = h_{AW} - h_{AZ} = S_W - S_Z = -4\widehat{s}_Z^2 \widehat{\epsilon}_2 / \alpha. \tag{10.61}$$

A heavy non-degenerate multiplet of fermions or scalars contributes positively to $T$ as

$$\rho_0 - 1 = \frac{1}{1 - \alpha T} - 1 \simeq \alpha T, \tag{10.62}$$

where $\rho_0$ is given in Eq. (10.58). The effects of non-standard Higgs representations cannot be separated from heavy non-degenerate multiplets unless the new physics has other consequences, such as vertex corrections. Most of the original papers defined $T$ to include the effects of loops only. However, we will redefine $T$ to include all new sources of SU(2) breaking, including non-standard Higgs, so that $T$ and $\rho_0$ are equivalent by Eq. (10.62).

A multiplet of heavy degenerate chiral fermions yields

$$S = C \sum_i \left(t_{3L}(i) - t_{3R}(i)\right)^2 / 3\pi ,  \quad (10.63)$$

where $t_{3L,R}(i)$ is the third component of weak isospin of the left-(right-)handed component of fermion $i$ and $C$ is the number of colors. For example, a heavy degenerate ordinary or mirror family would contribute $2/3\pi$ to $S$. In Technicolor models with QCD-like dynamics, one expects [155] $S \sim 0.45$ for an iso-doublet of techni-fermions, assuming $N_{TC} = 4$ techni-colors, while $S \sim 1.62$ for a full techni-generation with $N_{TC} = 4$; $T$ is harder to estimate because it is model dependent. In these examples one has $S \geq 0$. However, the QCD-like models are excluded on other grounds (flavor changing neutral-currents, and too-light quarks and pseudo-Goldstone bosons [162]). In particular, these estimates do not apply to models of walking Technicolor [162], for which $S$ can be smaller or even negative [163]. Other situations in which $S < 0$, such as loops involving scalars or Majorana particles, are also possible [164]. The simplest origin of $S < 0$ would probably be an additional heavy $Z'$ boson [152], which could mimic $S < 0$. Supersymmetric extensions of the SM generally give very small effects. See Refs. 115,165 and the Section on Supersymmetry in this *Review* for a complete set of references.

[115,165]. Most simple types of new physics yield $U = 0$, although there are counter-examples, such as the effects of anomalous triple gauge vertices [157].

The SM expressions for observables are replaced by

$$M_Z^2 = M_{Z0}^2 \frac{1 - \alpha T}{1 - G_F M_{Z0}^2 S / 2\sqrt{2}\pi} ,$$

$$M_W^2 = M_{W0}^2 \frac{1}{1 - G_F M_{W0}^2 (S + U) / 2\sqrt{2}\pi} , \quad (10.64)$$

where $M_{Z0}$ and $M_{W0}$ are the SM expressions (as functions of $m_t$ and $M_H$) in the $\overline{\text{MS}}$ scheme. Furthermore,

$$\Gamma_Z = \frac{1}{1 - \alpha T} M_Z^3 \beta_Z ,$$

$$\Gamma_W = M_W^3 \beta_W ,$$

$$A_i = \frac{1}{1 - \alpha T} A_{i0} , \quad (10.65)$$

where $\beta_Z$ and $\beta_W$ are the SM expressions for the reduced widths $\Gamma_{Z0}/M_{Z0}^3$ and $\Gamma_{W0}/M_{W0}^3$, $M_Z$ and $M_W$ are the physical masses, and $A_i$ ($A_{i0}$) is a neutral-current amplitude (in the SM).

The data allow a simultaneous determination of $\hat{s}_Z^2$ (from the Z-pole asymmetries), $S$ (from $M_Z$), $U$ (from $M_W$), $T$ (mainly from $\Gamma_Z$), $\alpha_s$ (from $R_\ell$, $\sigma_{\text{had}}$, and $\tau_\tau$), and $m_t$ (from CDF and DØ), with little correlation among the SM parameters:

$$\begin{aligned} S &= -0.13 \pm 0.10 \ (-0.08) \ , \\ T &= -0.17 \pm 0.12 \ (+0.09) \ , \\ U &= \phantom{-}0.22 \pm 0.13 \ (+0.01) \ , \end{aligned} \quad (10.66)$$

and $\hat{s}_Z^2 = 0.23119 \pm 0.00016$, $\alpha_s(M_Z) = 0.1222 \pm 0.0019$, $m_t = 177.2 \pm 4.2$ GeV, where the uncertainties are from the inputs. The central values assume $M_H = 117$ GeV, and in parentheses we show the change for $M_H = 300$ GeV. As can be seen, the SM parameters ($U$) can be determined with no (little) $M_H$ dependence. On the other hand, $S$, $T$, and $M_H$ cannot be obtained simultaneously, because the Higgs boson loops themselves are resembled approximately by oblique effects. Eqs. (10.66) show that negative (positive) contributions to the $S$ ($T$) parameter can weaken or entirely remove the strong constraints on $M_H$ from the SM fits. Specific models in which a large $M_H$ is compensated by new physics are reviewed in [166]. The parameters in Eqs. (10.66), which by definition are due to new physics only, all deviate by more than one standard deviation from the SM values of zero. However, these deviations are correlated. Fixing $U = 0$ (as is done in Fig. 10.3) will also move $S$ and $T$ to values compatible with zero within errors because the slightly high experimental value of $M_W$ favors a positive value for $S + U$. Using Eq. (10.62) the value of $\rho_0$ corresponding to $T$ is $0.9987 \pm 0.0009 \ (+0.0007)$. The values of the $\hat{\epsilon}$ parameters defined in Eq. (10.61) are

$$\begin{aligned} \hat{\epsilon}_3 &= -0.0011 \pm 0.0008 \ (-0.0006) \ , \\ \hat{\epsilon}_1 &= -0.0013 \pm 0.0009 \ (+0.0007) \ , \\ \hat{\epsilon}_2 &= -0.0019 \pm 0.0011 \ (-0.0001) \ . \end{aligned} \quad (10.67)$$

Unlike the original definition, we defined the quantities in Eqs. (10.67) to vanish identically in the absence of new physics and to correspond directly to the parameters $S$, $T$, and $U$ in Eqs. (10.66). There is a strong correlation (80%) between the $S$ and $T$ parameters. The allowed region in $S - T$ is shown in Fig. 10.3. From Eqs. (10.66) one obtains $S \leq 0.03 \ (-0.05)$ and $T \leq 0.02 \ (0.11)$ at 95% CL for $M_H = 117$ GeV (300 GeV). If one fixes $M_H = 600$ GeV and requires the constraint $S \geq 0$ (as is appropriate in QCD-like Technicolor models) then $S \leq 0.09$ (Bayesian) or $S \leq 0.06$ (frequentist). This rules out simple Technicolor models with many techni-doublets and QCD-like dynamics.

An extra generation of ordinary fermions is excluded at the 99.95% CL on the basis of the $S$ parameter alone, corresponding to $N_F = 2.92 \pm 0.27$ for the number of families. This result assumes that there are no new contributions to $T$ or $U$ and therefore that any new families are degenerate. In principle this restriction can be relaxed by allowing $T$ to vary as well, since $T > 0$ is expected from a non-degenerate extra family. However, the data currently favor $T < 0$, thus strengthening the exclusion limits. A more detailed analysis is required if the extra neutrino (or the extra down-type quark) is close to its direct mass limit [167]. This can drive $S$ to small or even negative values but at the expense of too-large contributions to $T$. These results are in agreement with a fit to the number of light neutrinos, $N_\nu = 2.986 \pm 0.007$ (which favors a larger value for $\alpha_s(M_Z) = 0.1228 \pm 0.0021$ mainly from $R_\ell$ and $\tau_\tau$). However, the $S$ parameter fits are valid even for a very heavy fourth family neutrino.

**Figure 10.3:** 1 $\sigma$ constraints (39.35%) on $S$ and $T$ from various inputs. $S$ and $T$ represent the contributions of new physics only. (Uncertainties from $m_t$ are included in the errors.) The contours assume $M_H = 117$ GeV except for the central and upper 90% CL contours allowed by all data, which are for $M_H = 340$ GeV and 1000 GeV, respectively. Data sets not involving $M_W$ are insensitive to $U$. Due to higher order effects, however, $U = 0$ has to be assumed in all fits. $\alpha_s$ is constrained using the $\tau$ lifetime as additional input in all fits.

There is no simple parametrization that is powerful enough to describe the effects of every type of new physics on every possible observable. The $S$, $T$, and $U$ formalism describes many types of heavy physics which affect only the gauge self-energies, and it can be applied to all precision observables. However, new physics which couples directly to ordinary fermions, such as heavy $Z'$ bosons [152] or mixing with exotic fermions [168] cannot be fully parametrized in the $S$, $T$, and $U$ framework. It is convenient to treat these types of new physics by parameterizations that are specialized to that particular class of theories (*e.g.*, extra $Z'$ bosons), or to consider specific models (which might contain, *e.g.*, $Z'$ bosons and exotic fermions with correlated parameters). Constraints on various types of new physics are reviewed in Refs. [43,91,169,170]. Fits to models with (extended) Technicolor and Supersymmetry are described, respectively, in Refs. [171], and [115,172]. The effects of compactified extra spatial dimensions at the TeV scale have been reviewed in [173], and constraints on Little Higgs models in [174].

An alternate formalism [175] defines parameters, $\epsilon_1$, $\epsilon_2$, $\epsilon_3$, $\epsilon_b$ in terms of the specific observables $M_W/M_Z$, $\Gamma_{\ell\ell}$, $A_{FB}^{(0,\ell)}$, and $R_b$. The definitions coincide with those for $\widehat{\epsilon}_i$ in Eqs. (10.60) and (10.61) for physics which affects gauge self-energies only, but the $\epsilon$'s now parametrize arbitrary types of new physics. However, the $\epsilon$'s are not related to other observables unless additional model-dependent assumptions are made. Another approach [176–178] parametrizes new physics in terms of gauge-invariant sets of operators. It is especially powerful in studying the effects of new physics on non-Abelian gauge vertices. The most general approach introduces deviation vectors [169]. Each type of new physics defines a deviation vector, the components of which are the deviations of each observable from its SM prediction, normalized to the experimental uncertainty. The length (direction) of the vector represents the strength (type) of new physics.

**Table 10.7:** 95% CL lower mass limits (in GeV) from low energy and $Z$ pole data on various extra $Z'$ gauge bosons, appearing in models of unification and string theory. $\rho_0$ free indicates a completely arbitrary Higgs sector, while $\rho_0 = 1$ restricts to Higgs doublets and singlets with still unspecified charges. The CDF bounds from searches for $\bar{p}p \to e^+e^-, \mu^+\mu^-$ [183] and the LEP 2 $e^+e^- \to f\bar{f}$ [41,184] bounds are listed in the last two columns, respectively. (The CDF bounds would be weakend if there are open supersymmetric or exotic decay channels.)

| Z' | $\rho_0$ free | $\rho_0 = 1$ | CDF (direct) | LEP 2 |
|---|---|---|---|---|
| $Z_\chi$ | 551 | 545 | 595 | 673 |
| $Z_\psi$ | 151 | 146 | 590 | 481 |
| $Z_\eta$ | 379 | 365 | 620 | 434 |
| $Z_{LR}$ | 570 | 564 | 630 | 804 |
| $Z_{SM}$ | 822 | 809 | 690 | 1787 |
| $Z_{\text{string}}$ | 582 | 578 | – | – |

One of the best motivated kinds of physics beyond the SM besides Supersymmetry are extra $Z'$ bosons. They do not spoil the observed approximate gauge coupling unification, and appear copiously in many Grand Unified Theories (GUTs), most Superstring models, as well as in dynamical symmetry breaking [171,179] and Little Higgs models [174]. For example, the SO(10) GUT contains an extra U(1) as can be seen from its maximal subgroup, SU(5) × U(1)$_\chi$. Similarly, the E$_6$ GUT contains the subgroup SO(10) × U(1)$_\psi$. The $Z_\psi$ possesses only axial-vector couplings to the ordinary fermions, and its mass is generally less constrained. The $Z_\eta$ boson is the linear combination $\sqrt{3/8}\, Z_\chi - \sqrt{5/8}\, Z_\psi$. The $Z_{LR}$ boson occurs in left-right models with gauge group SU(3)$_C$ × SU(2)$_L$ × SU(2)$_R$ × U(1)$_{B-L}$ ⊂ SO(10). The sequential $Z_{SM}$ boson is defined to have the same couplings to fermions as the SM $Z$ boson. Such a boson is not expected in the context of gauge theories unless it has different couplings to exotic fermions than the ordinary $Z$. However, it serves as a useful reference case when comparing constraints from various sources. It could also play the role of an excited state of the ordinary $Z$ in models with extra dimensions at the weak scale. Finally, we consider a Superstring motivated $Z_{string}$ boson appearing in a specific model [180]. The potential $Z'$ boson is in general a superposition of the SM $Z$ and the new boson associated with the extra U(1). The mixing angle $\theta$ satisfies,

$$\tan^2\theta = \frac{M_{Z_1^0}^2 - M_Z^2}{M_{Z'}^2 - M_{Z_1^0}^2},$$

where $M_{Z_1^0}$ is the SM value for $M_Z$ in the absence of mixing. Note, that $M_Z < M_{Z_1^0}$, and that the SM $Z$ couplings are changed by the mixing. If the Higgs U(1)′ quantum numbers are known, there will be an extra constraint,

$$\theta = C \frac{g_2}{g_1} \frac{M_Z^2}{M_{Z'}^2}, \qquad (10.68)$$

where $g_{1,2}$ are the U(1) and U(1)′ gauge couplings with $g_2 = \sqrt{\frac{5}{3}} \sin\theta_W \sqrt{\lambda} g_1$. $\lambda \sim 1$ (which we assume) if the GUT group breaks directly to SU(3) × SU(2) × U(1) × U(1)′. $C$ is a function of vacuum expectation values. For minimal Higgs sectors it can be found in reference [152]. Table 10.7 shows the 95% CL lower mass limits obtained from a somewhat earlier data set [181] for $\rho_0$ free and $\rho_0 = 1$, respectively. In cases of specific minimal Higgs sectors where $C$ is known, the $Z'$ mass limits are generally pushed into the TeV region. The limits on $|\theta|$ are typically $<$ few $\times 10^{-3}$. For more details see [181,182] and the Section on "The $Z'$ Searches" in this *Review*. Also listed in Table 10.7 are the direct lower limits on $Z'$ production from CDF [183] and LEP 2 bounds [41,184]. The final LEP 1 value for $\sigma_{had}$, some previous values for $Q_W(Cs)$, NuTeV, and $A_{FB}^{0,b}$ (for family-nonuniversal couplings [185]) modify the results and might even suggest the possible existence of a $Z'$ [144,186].

**References:**

1. S. Weinberg, Phys. Rev. Lett. **19**, 1264 (1967);

   A. Salam, p. 367 of *Elementary Particle Theory*, ed. N. Svartholm (Almquist and Wiksells, Stockholm, 1969);

   S.L. Glashow, J. Iliopoulos, and L. Maiani, Phys. Rev. **D2**, 1285 (1970).

2. H. Abele *et al.*, hep-ph/0312150.

3. For reviews, see G. Barbiellini and C. Santoni, Riv. Nuovo Cimento **9(2)**, 1 (1986);

   E.D. Commins and P.H. Bucksbaum, *Weak Interactions of Leptons and Quarks*, (Cambridge Univ. Press, Cambridge, 1983);

   W. Fetscher and H.J. Gerber, p. 657 of Ref. 4;

   J. Deutsch and P. Quin, p. 706 of Ref. 4;

   J.M. Conrad, M.H. Shaevitz, and T. Bolton, Rev. Mod. Phys. **70**, 1341 (1998).
4. *Precision Tests of the Standard Electroweak Model*, ed. P. Langacker (World Scientific, Singapore, 1995).
5. J. Erler and M. Luo, Phys. Lett. **B558**, 125 (2003).
6. CDF: T. Affolder *et al.*, Phys. Rev. **D63**, 032003 (2001).
7. E. Thomson for the CDF Collaboration, presented at the *31st SLAC Summer Institute* (SSI 2003, Menlo Park).
8. DØ: B. Abbott *et al.*, Phys. Rev. **D58**, 052001 (1998).
9. F. Canelli for the DØ Collaboration, presented at the *8th Conference on the Intersections of Particle and Nuclear Physics* (CIPANP 2003, New York).
10. L. Demortier *et al.*, preprint FERMILAB-TM-2084.
11. K. Melnikov and T. v. Ritbergen, Phys. Lett. **B482**, 99 (2000).
12. S.J. Brodsky, G.P. Lepage, and P.B. Mackenzie, Phys. Rev. **D28**, 228 (1983).
13. N. Gray *et al.*, Z. Phys. **C48**, 673 (1990).
14. For reviews, see the article on "The Higgs boson" in this *Review*;
   J. Gunion, H.E. Haber, G.L. Kane, and S. Dawson, *The Higgs Hunter's Guide*, (Addison-Wesley, Redwood City, 1990);
   M. Sher, Phys. Reports **179**, 273 (1989);
   M. Carena and H.E. Haber, Prog. Part. Nucl. Phys. **50**, 63 (2003).
15. P.J. Mohr and B.N. Taylor, Rev. Mod. Phys. **72**, 351 (2000).
16. TOPAZ: I. Levine *et al.*, Phys. Rev. Lett. **78**, 424 (1997);
   VENUS: S. Okada *et al.*, Phys. Rev. Lett. **81**, 2428 (1998);
   OPAL: G. Abbiendi *et al.*, Eur. Phys. J. **C13**, 553 (2000);
   L3: M. Acciarri *et al.*, Phys. Lett. **B476**, 40 (2000).
17. S. Fanchiotti, B. Kniehl, and A. Sirlin, Phys. Rev. **D48**, 307 (1993) and references therein.
18. J. Erler, Phys. Rev. **D59**, 054008 (1999).
19. CMD 2: R.R. Akhmetshin *et al.*, hep-ex/0308008.
20. M. Davier, S. Eidelman, A. Höcker, and Z. Zhang, hep-ph/0308213.
21. A.D. Martin and D. Zeppenfeld, Phys. Lett. **B345**, 558 (1995).
22. S. Eidelman and F. Jegerlehner, Z. Phys. **C67**, 585 (1995).
23. B.V. Geshkenbein and V.L. Morgunov, Phys. Lett. **B340**, 185 (1995) and Phys. Lett. **B352**, 456 (1995).
24. H. Burkhardt and B. Pietrzyk, Phys. Lett. **B356**, 398 (1995).
25. M.L. Swartz, Phys. Rev. **D53**, 5268 (1996).
26. R. Alemany, M. Davier, and A. Höcker, Eur. Phys. J. **C2**, 123 (1998).
27. N.V. Krasnikov and R. Rodenberg, Nuovo Cimento **111A**, 217 (1998).

28. M. Davier and A. Höcker, Phys. Lett. **B419**, 419 (1998).
29. J.H. Kühn and M. Steinhauser, Phys. Lett. **B437**, 425 (1998).
30. M. Davier and A. Höcker, Phys. Lett. **B435**, 427 (1998).
31. S. Groote, J.G. Körner, K. Schilcher, N.F. Nasrallah, Phys. Lett. **B440**, 375 (1998).
32. A.D. Martin, J. Outhwaite, and M.G. Ryskin, Phys. Lett. **B492**, 69 (2000).
33. H. Burkhardt and B. Pietrzyk, Phys. Lett. **B513**, 46 (2001).
34. J.F. de Troconiz and F.J. Yndurain, Phys. Rev. **D65**, 093002 (2002).
35. F. Jegerlehner, `hep-ph/0308117`.
36. BES: J.Z. Bai *et al.*, Phys. Rev. Lett. **88**, 101802 (2002);
    G.S. Huang, `hep-ex/0105074`.
37. S. Binner, J.H. Kühn, and K. Melnikov, Phys. Lett. **B459**, 279 (1999).
38. KLOE: A. Aloisio *et al.*, `hep-ex/0307051`.
39. W.J. Marciano and A. Sirlin, Phys. Rev. Lett. **61**, 1815 (1988).
40. T. van Ritbergen and R.G. Stuart, Phys. Rev. Lett. **82**, 488 (1999).
41. ALEPH, DELPHI, L3, OPAL, LEP Electroweak Working Group, and SLD Heavy Flavor Group, `hep-ex/0312023`.
42. Earlier analyses include U. Amaldi *et al.*, Phys. Rev. **D36**, 1385 (1987);
    G. Costa *et al.*, Nucl. Phys. **B297**, 244 (1988);
    Deep inelastic scattering is considered by G.L. Fogli and D. Haidt, Z. Phys. **C40**, 379 (1988);
    P. Langacker and M. Luo, Phys. Rev. **D44**, 817 (1991);
    For more recent analyses, see Ref. 43.
43. P. Langacker, p. 883 of Ref. 4;
    J. Erler and P. Langacker, Phys. Rev. **D52**, 441 (1995);
    Neutrino scattering is reviewed by J.M. Conrad *et al.* in Ref. 3;
    Nonstandard neutrino interactions are surveyed in Z. Berezhiani and A. Rossi, Phys. Lett. **B535**, 207 (2002);
    S. Davidson, C. Pena-Garay, N. Rius, and A. Santamaria, JHEP **0303**, 011 (2003).
44. A. Sirlin, Phys. Rev. **D22**, 971 (1980) and *ibid.* **29**, 89 (1984);
    D.C. Kennedy *et al.*, Nucl. Phys. **B321**, 83 (1989);
    D.C. Kennedy and B.W. Lynn, Nucl. Phys. **B322**, 1 (1989);
    D.Yu. Bardin *et al.*, Z. Phys. **C44**, 493 (1989);
    W. Hollik, Fortsch. Phys. **38**, 165 (1990);
    For reviews, see the articles by W. Hollik, pp. 37 and 117, and W. Marciano, p. 170 in Ref. 4. Extensive references to other papers are given in Ref. 42.

45. W. Hollik in Ref. 44 and references therein;
    V.A. Novikov, L.B. Okun, and M.I. Vysotsky, Nucl. Phys. **B397**, 35 (1993).
46. W.J. Marciano and J.L. Rosner, Phys. Rev. Lett. **65**, 2963 (1990).
47. G. Degrassi, S. Fanchiotti, and A. Sirlin, Nucl. Phys. **B351**, 49 (1991).
48. G. Degrassi and A. Sirlin, Nucl. Phys. **B352**, 342 (1991).
49. P. Gambino and A. Sirlin, Phys. Rev. **D49**, 1160 (1994).
50. ZFITTER: D. Bardin *et al.*, Comput. Phys. Commun. **133**, 229 (2001) and references therein.
51. R. Barbieri *et al.*, Phys. Lett. **B288**, 95 (1992);
    R. Barbieri *et al.*, Nucl. Phys. **B409**, 105 (1993).
52. J. Fleischer, O.V. Tarasov, and F. Jegerlehner, Phys. Lett. **B319**, 249 (1993).
53. G. Degrassi, P. Gambino, and A. Vicini, Phys. Lett. **B383**, 219 (1996);
    G. Degrassi, P. Gambino, and A. Sirlin, Phys. Lett. **B394**, 188 (1997).
54. A. Freitas, W. Hollik, W. Walter, and G. Weiglein, Phys. Lett. **B495**, 338 (2000) and Nucl. Phys. **B632**, 189 (2002).
55. M. Awramik and M. Czakon, Phys. Rev. Lett. **89**, 241801 (2002);
    A. Onishchenko and O. Veretin, Phys. Lett. **B551**, 111 (2003).
56. A. Djouadi and C. Verzegnassi, Phys. Lett. **B195**, 265 (1987);
    A. Djouadi, Nuovo Cimento **100A**, 357 (1988);
    B.A. Kniehl, Nucl. Phys. **B347**, 86 (1990);
    A. Djouadi and P. Gambino, Phys. Rev. **D49**, 3499 (1994), **D49**, 4705 (1994), and **D53**, 4111(E) (1996).
57. K.G. Chetyrkin, J.H. Kühn, and M. Steinhauser, Phys. Lett. **B351**, 331 (1995);
    L. Avdeev *et al.*, Phys. Lett. **B336**, 560 (1994) and **B349**, 597(E) (1995).
58. J. Fleischer *et al.*, Phys. Lett. **B293**, 437 (1992);
    K.G. Chetyrkin, A. Kwiatkowski, and M. Steinhauser, Mod. Phys. Lett. **A8**, 2785 (1993).
59. A. Czarnecki and J.H. Kühn, Phys. Rev. Lett. **77**, 3955 (1996).
60. R. Harlander, T. Seidensticker, and M. Steinhauser, Phys. Lett. **B426**, 125 (1998);
    J. Fleischer *et al.*, Phys. Lett. **B459**, 625 (1999).
61. J. Erler, `hep-ph/0005084`.
62. For reviews, see F. Perrier, p. 385 of Ref. 4;
    J.M. Conrad *et al.* in Ref. 3.
63. CDHS: H. Abramowicz *et al.*, Phys. Rev. Lett. **57**, 298 (1986);
    CDHS: A. Blondel *et al.*, Z. Phys. **C45**, 361 (1990).

64. CHARM: J.V. Allaby *et al.*, Phys. Lett. **B177**, 446 (1986);
    CHARM: J.V. Allaby *et al.*, Z. Phys. **C36**, 611 (1987).
65. CCFR: C.G. Arroyo *et al.*, Phys. Rev. Lett. **72**, 3452 (1994);
    CCFR: K.S. McFarland *et al.*, Eur. Phys. J. **C1**, 509 (1998).
66. R.M. Barnett, Phys. Rev. **D14**, 70 (1976);
    H. Georgi and H.D. Politzer, Phys. Rev. **D14**, 1829 (1976).
67. LAB-E: S.A. Rabinowitz *et al.*, Phys. Rev. Lett. **70**, 134 (1993).
68. E.A. Paschos and L. Wolfenstein, Phys. Rev. **D7**, 91 (1973).
69. NuTeV: G. P. Zeller *et al.*, Phys. Rev. Lett. **88**, 091802 (2002).
70. S. Davidson *et al.*, JHEP **0202**, 037 (2002).
71. NuTeV: M. Goncharov *et al.*, Phys. Rev. **D64**, 112006 (2001).
72. NuTeV: G.P. Zeller *et al.*, Phys. Rev. **D65**, 111103 (2002);
    NuTeV: R. H. Bernstein *et al.*, J. Phys. G **29**, 1919 (2003).
73. K.P.O. Diener, S. Dittmaier, and W. Hollik, `hep-ph/0310364`.
74. CHARM: J. Dorenbosch *et al.*, Z. Phys. **C41**, 567 (1989).
75. CALO: L.A. Ahrens *et al.*, Phys. Rev. **D41**, 3297 (1990).
76. CHARM II: P. Vilain *et al.*, Phys. Lett. **B335**, 246 (1994).
77. See also J. Panman, p. 504 of Ref. 4.
78. ILM: R.C. Allen *et al.*, Phys. Rev. **D47**, 11 (1993);
    LSND: L.B. Auerbach *et al.*, Phys. Rev. **D63**, 112001 (2001).
79. SSF: C.Y. Prescott *et al.*, Phys. Lett. **B84**, 524 (1979);
    For a review, see P. Souder, p. 599 of Ref. 4.
80. For reviews and references to earlier work, see M.A. Bouchiat and L. Pottier, Science **234**, 1203 (1986);
    B.P. Masterson and C.E. Wieman, p. 545 of Ref. 4.
81. Cesium (Boulder): C.S. Wood *et al.*, Science **275**, 1759 (1997).
82. Thallium (Oxford): N.H. Edwards *et al.*, Phys. Rev. Lett. **74**, 2654 (1995);
    Thallium (Seattle): P.A. Vetter *et al.*, Phys. Rev. Lett. **74**, 2658 (1995).
83. Lead (Seattle): D.M. Meekhof *et al.*, Phys. Rev. Lett. **71**, 3442 (1993).
84. Bismuth (Oxford): M.J.D. MacPherson *et al.*, Phys. Rev. Lett. **67**, 2784 (1991).
85. V.A. Dzuba, V.V. Flambaum, and O.P. Sushkov, Phys. Lett. **141A**, 147 (1989);
    S.A. Blundell, J. Sapirstein, and W.R. Johnson, Phys. Rev. Lett. **65**, 1411 (1990) and Phys. Rev. **D45**, 1602 (1992);
    For a review, see S.A. Blundell, W.R. Johnson, and J. Sapirstein, p. 577 of Ref. 4.
86. S.C. Bennett and C.E. Wieman, Phys. Rev. Lett. **82**, 2484 (1999).

87. V.A. Dzuba, V.V. Flambaum, and O.P. Sushkov, Phys. Rev. **A56**, R4357 (1997).

88. M.A. Bouchiat and J. Guéna, J. Phys. (France) **49**, 2037 (1988).

89. A. Derevianko, Phys. Rev. Lett. **85**, 1618 (2000);

    V.A. Dzuba, C. Harabati, and W.R. Johnson, Phys. Rev. **A63**, 044103 (2001);

    M.G. Kozlov, S.G. Porsev, and I.I. Tupitsyn, Phys. Rev. Lett. **86**, 3260 (2001).

90. A.I. Milstein and O.P. Sushkov, Phys. Rev. **A66**, 022108 (2002);

    W.R. Johnson, I. Bednyakov, and G. Soff, Phys. Rev. Lett. **87**, 233001 (2001);

    V.A. Dzuba, V.V. Flambaum, and J.S. Ginges, Phys. Rev. **D66**, 076013 (2002);

    M.Y. Kuchiev and V.V. Flambaum, Phys. Rev. Lett. **89**, 283002 (2002);

    A.I. Milstein, O.P. Sushkov, and I.S. Terekhov, Phys. Rev. Lett. **89**, 283003 (2002);

    For a recent review, see J.S.M. Ginges and V.V. Flambaum, `physics/0309054`.

91. J. Erler, A. Kurylov, and M.J. Ramsey-Musolf, Phys. Rev. **D68**, 016006 (2003).

92. V.A. Dzuba *et al.*, J. Phys. **B20**, 3297 (1987).

93. Ya.B. Zel'dovich, Sov. Phys. JETP **6**, 1184 (1958);

    For recent discussions, see V.V. Flambaum and D.W. Murray, Phys. Rev. **C56**, 1641 (1997);

    W.C. Haxton and C.E. Wieman, Ann. Rev. Nucl. Part. Sci. **51**, 261 (2001).

94. J.L. Rosner, Phys. Rev. **D53**, 2724 (1996).

95. S.J. Pollock, E.N. Fortson, and L. Wilets, Phys. Rev. **C46**, 2587 (1992);

    B.Q. Chen and P. Vogel, Phys. Rev. **C48**, 1392 (1993).

96. B.W. Lynn and R.G. Stuart, Nucl. Phys. **B253**, 216 (1985).

97. *Physics at LEP*, ed. J. Ellis and R. Peccei, CERN 86–02, Vol. 1.

98. C. Kiesling, *Tests of the Standard Theory of Electroweak Interactions*, (Springer-Verlag, New York, 1988);

    R. Marshall, Z. Phys. **C43**, 607 (1989);

    Y. Mori *et al.*, Phys. Lett. **B218**, 499 (1989);

    D. Haidt, p. 203 of Ref. 4.

99. For reviews, see D. Schaile, p. 215, and A. Blondel, p. 277 of Ref. 4.

100. M. Elsing, presented at the *International Europhysics Conference on High Energy Physics* (EPS 2003, Aachen).

101. SLD: K. Abe *et al.*, Phys. Rev. Lett. **84**, 5945 (2000).

102. SLD: K. Abe *et al.*, Phys. Rev. Lett. **85**, 5059 (2000).
103. SLD: K. Abe *et al.*, Phys. Rev. Lett. **86**, 1162 (2001).
104. DELPHI: P. Abreu *et al.*, Z. Phys. **C67**, 1 (1995);
    OPAL: K. Ackerstaff *et al.*, Z. Phys. **C76**, 387 (1997).
105. SLD: K. Abe *et al.*, Phys. Rev. Lett. **78**, 17 (1997).
106. ALEPH, DELPHI, L3, and OPAL Collaborations, and the LEP Working Group for Higgs Boson Searches: D. Abbaneo *et al.*, Phys. Lett. **B565**, 61 (2003).
107. A. Leike, T. Riemann, and J. Rose, Phys. Lett. **B273**, 513 (1991);
    T. Riemann, Phys. Lett. **B293**, 451 (1992).
108. E158: P.L. Anthony *et al.*, `hep-ex/0312035`;
    the implications are discussed in A. Czarnecki and W.J. Marciano, Int. J. Mod. Phys. A **15**, 2365 (2000).
109. G. S. Mitchell, `hep-ex/0308049`;
    the implications are discussed in Ref. 91.
110. Belle: K. Abe *et al.*, Phys. Lett. **B511**, 151 (2001).
111. CLEO: S. Chen *et al.*, Phys. Rev. Lett. **87**, 251807 (2001).
112. BaBar: B. Aubert *et al.*, `hep-ex/0207076`.
113. A. Ali and C. Greub, Phys. Lett. **B259**, 182 (1991);
    A.L. Kagan and M. Neubert, Eur. Phys. J. **C7**, 5 (1999).
114. A. Czarnecki and W.J. Marciano, Phys. Rev. Lett. **81**, 277 (1998).
115. J. Erler and D.M. Pierce, Nucl. Phys. **B526**, 53 (1998).
116. Y. Nir, Phys. Lett. **B221**, 184 (1989);
    K. Adel and Y.P. Yao, Phys. Rev. **D49**, 4945 (1994);
    C. Greub, T. Hurth, and D. Wyler, Phys. Rev. **D54**, 3350 (1996);
    K.G. Chetyrkin, M. Misiak, and M. Münz, Phys. Lett. **B400**, 206 (1997);
    C. Greub and T. Hurth, Phys. Rev. **D56**, 2934 (1997);
    M. Ciuchini *et al.*, Nucl. Phys. **B527**, 21 (1998) and **B534**, 3 (1998);
    F.M. Borzumati and C. Greub, Phys. Rev. **D58**, 074004 (1998) and **D59**, 057501 (1999);
    A. Strumia, Nucl. Phys. **B532**, 28 (1998).
117. E821: H.N. Brown *et al.*, Phys. Rev. Lett. **86**, 2227 (2001);
    E821: G.W. Bennett, *et al.*, Phys. Rev. Lett. **89**, 101804 (2002).
118. For reviews, see V.W. Hughes and T. Kinoshita, Rev. Mod. Phys. **71**, S133 (1999);
    A. Czarnecki and W.J. Marciano, Phys. Rev. **D64**, 013014 (2001);
    T. Kinoshita, J. Phys. **G29**, 9 (2003).
119. S.J. Brodsky and J.D. Sullivan, Phys. Rev. **D156**, 1644 (1967);
    T. Burnett and M.J. Levine, Phys. Lett. **B24**, 467 (1967);
    R. Jackiw and S. Weinberg, Phys. Rev. **D5**, 2473 (1972);

I. Bars and M. Yoshimura, Phys. Rev. **D6**, 374 (1972);

K. Fujikawa, B.W. Lee, and A.I. Sanda, Phys. Rev. **D6**, 2923 (1972);

G. Altarelli, N. Cabibbo, and L. Maiani, Phys. Lett. **B40**, 415 (1972);

W.A. Bardeen, R. Gastmans, and B.E. Laurup, Nucl. Phys. **B46**, 315 (1972).

120. T.V. Kukhto, E.A. Kuraev, A. Schiller, and Z.K. Silagadze, Nucl. Phys. **B371**, 567 (1992);

    S. Peris, M. Perrottet, and E. de Rafael, Phys. Lett. **B355**, 523 (1995);

    A. Czarnecki, B. Krause, and W.J. Marciano, Phys. Rev. **D52**, 2619 (1995) and Phys. Rev. Lett. **76**, 3267 (1996).

121. G. Degrassi and G. Giudice, Phys. Rev. **D58**, 053007 (1998).

122. F. Matorras (DELPHI), contributed paper to the *International Europhysics Conference on High Energy Physics* (EPS 2003, Aachen).

123. V. Cirigliano, G. Ecker and H. Neufeld, JHEP **0208**, 002 (2002).

124. J. Erler, `hep-ph/0211345`.

125. M. Knecht and A. Nyffeler, Phys. Rev. **D65**, 073034 (2002).

126. M. Hayakawa and T. Kinoshita, `hep-ph/0112102`;

    J. Bijnens, E. Pallante and J. Prades, Nucl. Phys. **B626**, 410 (2002).

127. B. Krause, Phys. Lett. **B390**, 392 (1997).

128. J. Erler and M. Luo, Phys. Rev. Lett. **87**, 071804 (2001).

129. J.L. Lopez, D.V. Nanopoulos, and X. Wang, Phys. Rev. **D49**, 366 (1994);

    for recent reviews, see Ref. 118.

130. E821: G.W. Bennett *et al.*, `hep-ex/0401008`.

131. S. Ghozzi and F. Jegerlehner, `hep-ph/0310181`.

132. K. Hagiwara, A.D. Martin, D. Nomura, and T. Teubner, `hep-ph/0312250`.

133. A comprehensive report and further references can be found in K.G. Chetyrkin, J.H. Kühn, and A. Kwiatkowski, Phys. Reports **277**, 189 (1996).

134. J. Schwinger, *Particles, Sources and Fields*, Vol. II, (Addison-Wesley, New York, 1973);

    K.G. Chetyrkin, A.L. Kataev, and F.V. Tkachev, Phys. Lett. **B85**, 277 (1979);

    M. Dine and J. Sapirstein, Phys. Rev. Lett. **43**, 668 (1979);

    W. Celmaster, R.J. Gonsalves, Phys. Rev. Lett. **44**, 560 (1980);

    S.G. Gorishnii, A.L. Kataev, and S.A. Larin, Phys. Lett. **B212**, 238 (1988) and **B259**, 144 (1991);

L.R. Surguladze and M.A. Samuel, Phys. Rev. Lett. **66**, 560 (1991) and 2416(E).

135. W. Bernreuther and W. Wetzel, Z. Phys. **11**, 113 (1981) and Phys. Rev. **D24**, 2724 (1982);

    B.A. Kniehl, Phys. Lett. **B237**, 127 (1990);

    K.G. Chetyrkin, Phys. Lett. **B307**, 169 (1993);

    A.H. Hoang *et al.*, Phys. Lett. **B338**, 330 (1994);

    S.A. Larin, T. van Ritbergen, and J.A.M. Vermaseren, Nucl. Phys. **B438**, 278 (1995).

136. T.H. Chang, K.J.F. Gaemers, and W.L. van Neerven, Nucl. Phys. **B202**, 407 (1980);

    J. Jersak, E. Laermann, and P.M. Zerwas, Phys. Lett. **B98**, 363 (1981) and Phys. Rev. **D25**, 1218 (1982);

    S.G. Gorishnii, A.L. Kataev, and S.A. Larin, Nuovo Cimento **92**, 117 (1986);

    K.G. Chetyrkin and J.H. Kühn, Phys. Lett. **B248**, 359 (1990) and *ibid.* **406**, 102 (1997);

    K.G. Chetyrkin, J.H. Kühn, and A. Kwiatkowski, Phys. Lett. **B282**, 221 (1992).

137. B.A. Kniehl and J.H. Kühn, Phys. Lett. **B224**, 229 (1990) and Nucl. Phys. **B329**, 547 (1990);

    K.G. Chetyrkin and A. Kwiatkowski, Phys. Lett. **B305**, 285 (1993) and **B319**, 307 (1993);

    S.A. Larin, T. van Ritbergen, and J.A.M. Vermaseren, Phys. Lett. **B320**, 159 (1994);

    K.G. Chetyrkin and O.V. Tarasov, Phys. Lett. **B327**, 114 (1994).

138. A.L. Kataev, Phys. Lett. **B287**, 209 (1992).

139. D. Albert *et al.*, Nucl. Phys. **B166**, 460 (1980);

    F. Jegerlehner, Z. Phys. **C32**, 425 (1986);

    A. Djouadi, J.H. Kühn, and P.M. Zerwas, Z. Phys. **C46**, 411 (1990);

    A. Borrelli *et al.*, Nucl. Phys. **B333**, 357 (1990).

140. A.A. Akhundov, D.Yu. Bardin, and T. Riemann, Nucl. Phys. **B276**, 1 (1986);

    W. Beenakker and W. Hollik, Z. Phys. **C40**, 141 (1988);

    B.W. Lynn and R.G. Stuart, Phys. Lett. **B352**, 676 (1990);

    J. Bernabeu, A. Pich, and A. Santamaria, Nucl. Phys. **B363**, 326 (1991).

141. UA2: S. Alitti *et al.*, Phys. Lett. **B276**, 354 (1992);

    CDF: T. Affolder *et al.*, Phys. Rev. **D64**, 052001 (2001);

    DØ: V. M. Abazov *et al.*, Phys. Rev. **D66**, 012001 (2002);

    CDF and DØ Collaborations: `hep-ex/0311039`.

142. J. Erler, J.L. Feng, and N. Polonsky, Phys. Rev. Lett. **78**, 3063 (1997).

143. D. Choudhury, T.M.P. Tait and C.E.M. Wagner, Phys. Rev. **D65**, 053002 (2002).

144. J. Erler and P. Langacker, Phys. Rev. Lett. **84**, 212 (2000).

145. DELPHI: P. Abreu *et al.*, Z. Phys. **C**, 70 (1996);

    DELPHI: P. Abreu *et al.*, in the Proceedings of the *International Europhysics Conference on High Energy Physics* (Jerusalem, 1997).

146. P. Schleper, presented at the *International Europhysics Conference on High Energy Physics* (EPS 2003, Aachen).

147. HPQCD: C.T. Davies *et al.*, `hep-lat/0304004`.

148. J. Erler, Phys. Rev. **D63**, 071301 (2001).

149. P. Langacker and N. Polonsky, Phys. Rev. **D52**, 3081 (1995);

    J. Bagger, K.T. Matchev, and D. Pierce, Phys. Lett. **B348**, 443 (1995).

150. A.L. Kataev and V.V. Starshenko, Mod. Phys. Lett. **A10**, 235 (1995).

151. M. Veltman, Nucl. Phys. **B123**, 89 (1977);

    M. Chanowitz, M.A. Furman, and I. Hinchliffe, Phys. Lett. **B78**, 285 (1978).

152. P. Langacker and M. Luo, Phys. Rev. **D45**, 278 (1992) and references therein.

153. A. Denner, R.J. Guth, and J.H. Kühn, Phys. Lett. **B240**, 438 (1990).

154. S. Bertolini and A. Sirlin, Phys. Lett. **B257**, 179 (1991).

155. M. Peskin and T. Takeuchi, Phys. Rev. Lett. **65**, 964 (1990) and Phys. Rev. **D46**, 381 (1992);

    M. Golden and L. Randall, Nucl. Phys. **B361**, 3 (1991).

156. D. Kennedy and P. Langacker, Phys. Rev. Lett. **65**, 2967 (1990) and Phys. Rev. **D44**, 1591 (1991).

157. G. Altarelli and R. Barbieri, Phys. Lett. **B253**, 161 (1990).

158. B. Holdom and J. Terning, Phys. Lett. **B247**, 88 (1990).

159. B.W. Lynn, M.E. Peskin, and R.G. Stuart, p. 90 of Ref. 97.

160. An alternative formulation is given by K. Hagiwara *et al.*, Z. Phys. **C64**, 559 (1994) and *ibid.* **C68**, 352(E) (1995);

    K. Hagiwara, D. Haidt, and S. Matsumoto, Eur. Phys. J. **C2**, 95 (1998).

161. I. Maksymyk, C.P. Burgess, and D. London, Phys. Rev. **D50**, 529 (1994);

    C.P. Burgess *et al.*, Phys. Lett. **B326**, 276 (1994).

162. K. Lane, in the Proceedings of the *27th International Conference on High Energy Physics* (Glasgow, 1994).

163. E. Gates and J. Terning, Phys. Rev. Lett. **67**, 1840 (1991);

    R. Sundrum and S.D.H. Hsu, Nucl. Phys. **B391**, 127 (1993);

R. Sundrum, Nucl. Phys. **B395**, 60 (1993);

M. Luty and R. Sundrum, Phys. Rev. Lett. **70**, 529 (1993);

T. Appelquist and J. Terning, Phys. Lett. **B315**, 139 (1993).

164. H. Georgi, Nucl. Phys. **B363**, 301 (1991);

    M.J. Dugan and L. Randall, Phys. Lett. **B264**, 154 (1991).

165. R. Barbieri *et al.*, Nucl. Phys. **B341**, 309 (1990).

166. M.E. Peskin and J.D. Wells, Phys. Rev. **D64**, 093003 (2001).

167. H.J. He, N. Polonsky, and S. Su, Phys. Rev. **D64**, 053004 (2001);

    V.A. Novikov, L.B. Okun, A.N. Rozanov, and M.I. Vysotsky, Sov. Phys. JETP **76**, 127 (2002) and references therein.

168. For a review, see D. London, p. 951 of Ref. 4;

    a recent analysis is M.B. Popovic and E.H. Simmons, Phys. Rev. **D58**, 095007 (1998);

    for collider implications, see T.C. Andre and J.L. Rosner, `hep-ph/0309254`.

169. P. Langacker, M. Luo, and A.K. Mann, Rev. Mod. Phys. **64**, 87 (1992);

    M. Luo, p. 977 of Ref. 4.

170. F.S. Merritt *et al.*, p. 19 of *Particle Physics: Perspectives and Opportunities: Report of the DPF Committee on Long Term Planning*, ed. R. Peccei *et al.* (World Scientific, Singapore, 1995).

171. C.T. Hill and E.H. Simmons, Phys. Reports **381**, 235 (2003).

172. G. Altarelli *et al.*, JHEP **0106**, 018 (2001);

    A. Kurylov, M.J. Ramsey-Musolf, and S. Su, Nucl. Phys. **B667**, 321 (2003) and Phys. Rev. **D68**, 035008 (2003);

    W. de Boer and C. Sander, `hep-ph/0307049`;

    S. Heinemeyer and G. Weiglein, `hep-ph/0307177`;

    J.R. Ellis, K.A. Olive, Y. Santoso, and V.C. Spanos, `hep-ph/0310356`.

173. K. Agashe, A. Delgado, M.J. May, and R. Sundrum, JHEP **0308**, 050 (2003);

    M. Carena *et al.*, Phys. Rev. **D68**, 035010 (2003);

    for reviews, see the articles on "Extra Dimensions" in this *Review* and I. Antoniadis, `hep-th/0102202`.

174. J.L. Hewett, F.J. Petriello, and T.G. Rizzo, JHEP **0310**, 062 (2003);

    C. Csaki *et al.*, Phys. Rev. **D67**, 115002 (2003) and *ibid.* **68**, 035009 (2003);

    T. Gregoire, D.R. Smith, and J.G. Wacker, `hep-ph/0305275`;

    M. Perelstein, M.E. Peskin, and A. Pierce, `hep-ph/0310039`;

    R. Casalbuoni, A. Deandrea, and M. Oertel, `hep-ph/0311038`;

    W. Kilian and J. Reuter, `hep-ph/0311095`.

175. G. Altarelli, R. Barbieri, and S. Jadach, Nucl. Phys. **B369**, 3 (1992) and **B376**, 444(E) (1992).
176. A. De Rújula *et al.*, Nucl. Phys. **B384**, 3 (1992).
177. K. Hagiwara *et al.*, Phys. Rev. **D48**, 2182 (1993).
178. C.P. Burgess and D. London, Phys. Rev. **D48**, 4337 (1993).
179. R.S. Chivukula and E.H. Simmons, Phys. Rev. **D66**, 015006 (2002).
180. S. Chaudhuri *et al.*, Nucl. Phys. **B456**, 89 (1995);
    G. Cleaver *et al.*, Phys. Rev. **D59**, 055005 (1999).
181. J. Erler and P. Langacker, Phys. Lett. **B456**, 68 (1999).
182. T. Appelquist, B.A. Dobrescu, and A.R. Hopper, Phys. Rev. **D68**, 035012 (2003);
    R.S. Chivukula, H.J. He, J. Howard, and E.H. Simmons, `hep-ph/0307209`.
183. CDF: F. Abe *et al.*, Phys. Rev. Lett. **79**, 2192 (1997).
184. K.M. Cheung, Phys. Lett. **B517**, 167 (2001).
185. P. Langacker and M. Plümacher, Phys. Rev. **D62**, 013006 (2000).
186. R. Casalbuoni, S. De Curtis, D. Dominici, and R. Gatto, Phys. Lett. **B460**, 135 (1999);
    J.L. Rosner, Phys. Rev. **D61**, 016006 (2000).

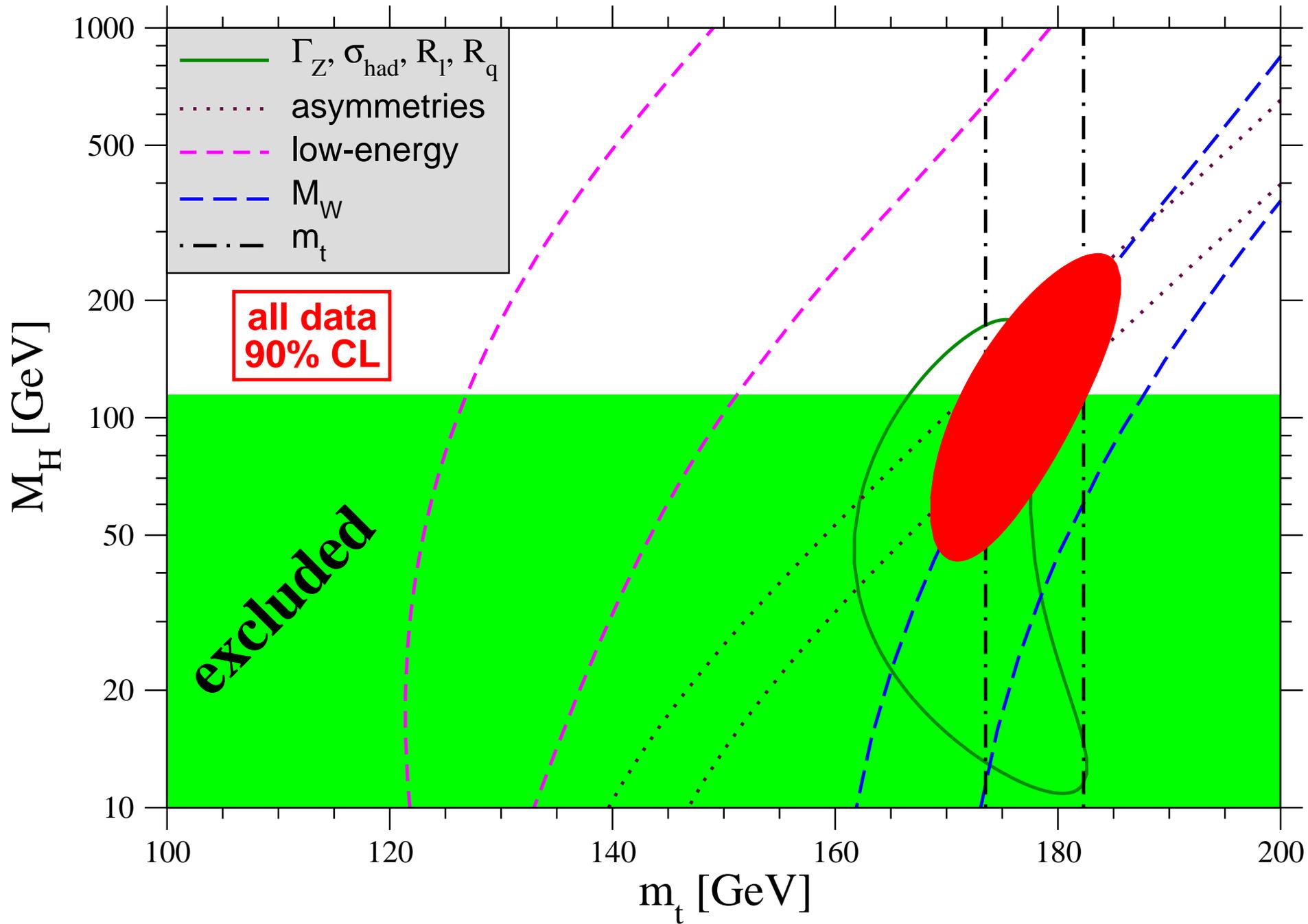

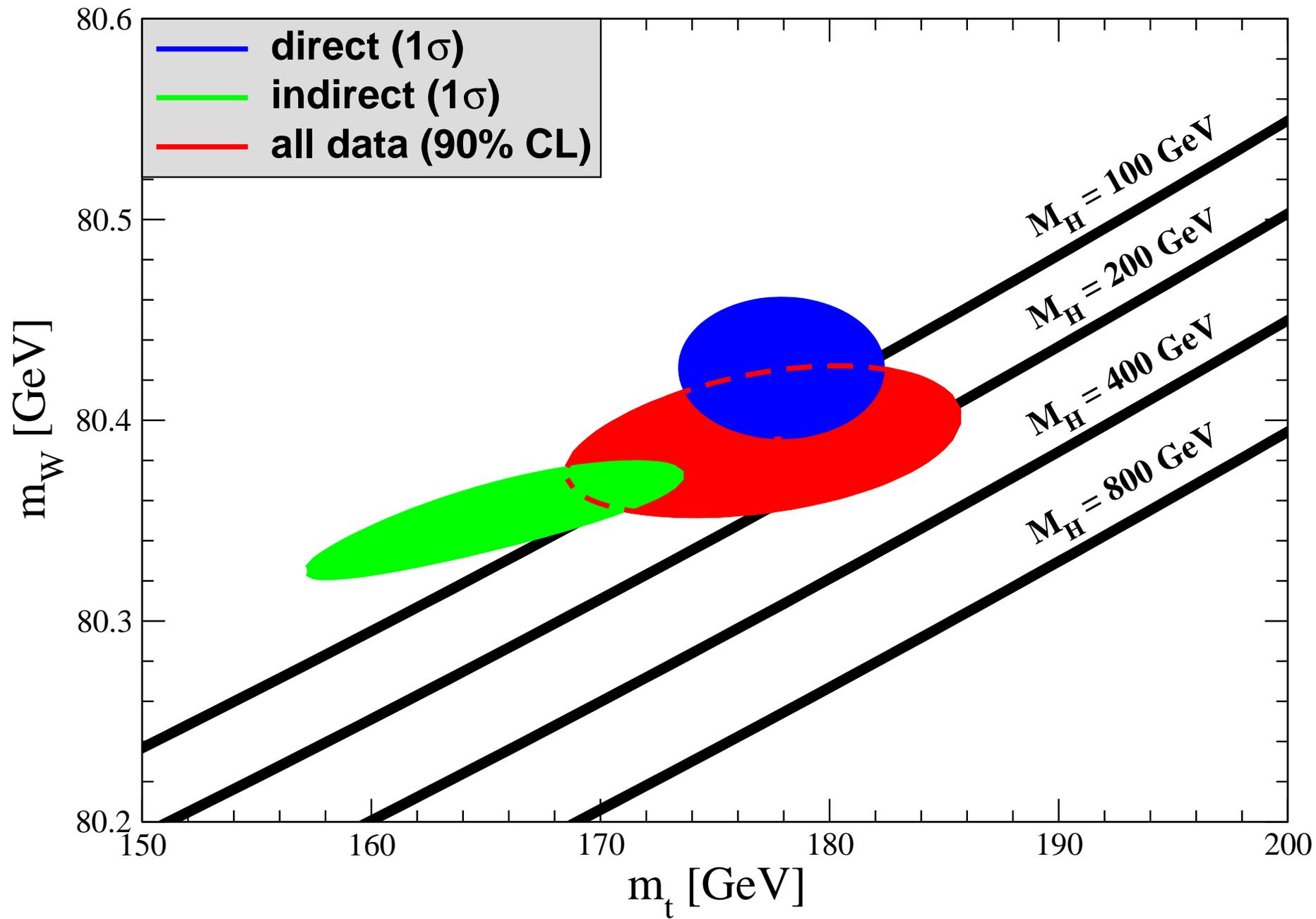

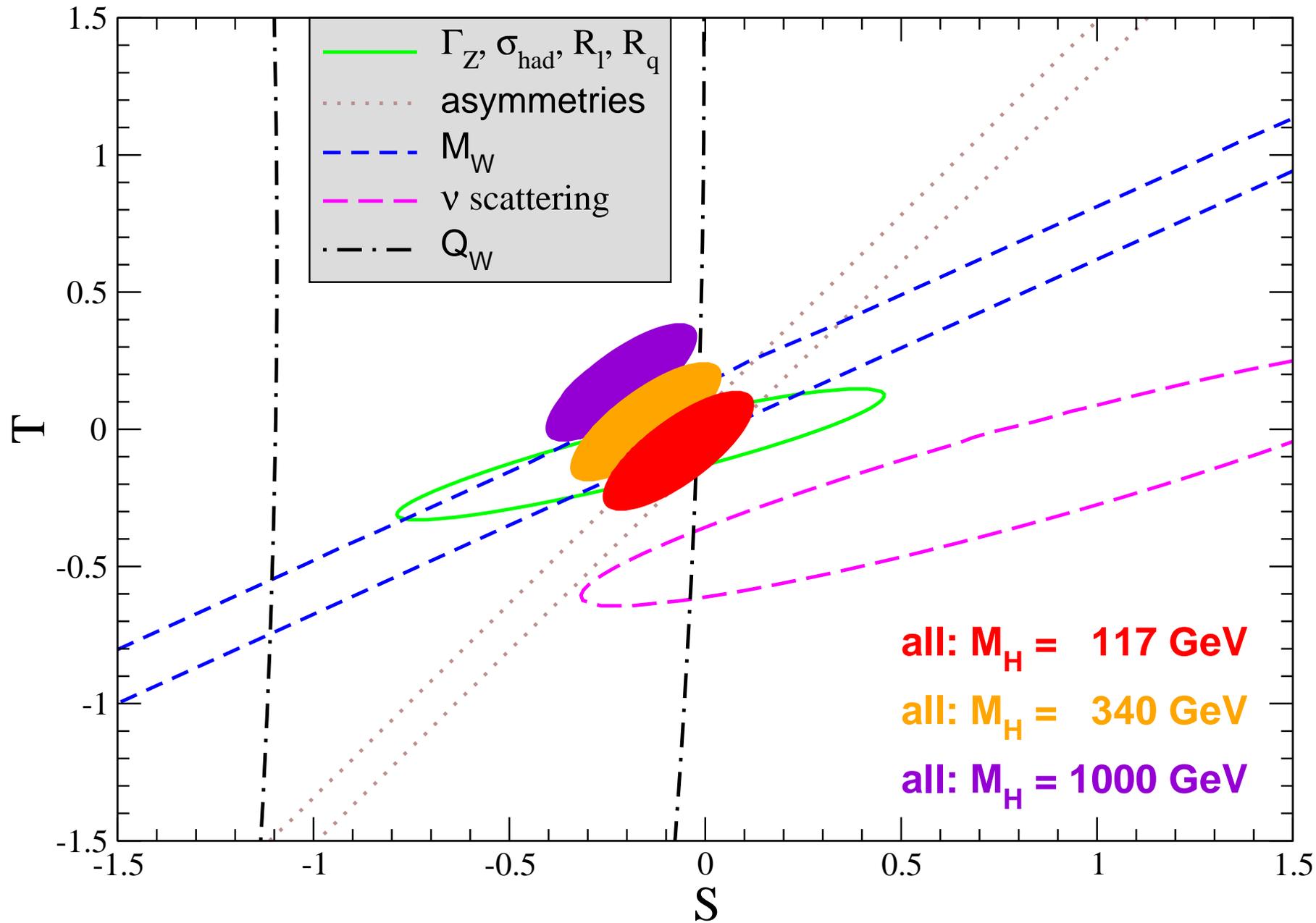